\documentclass[aps,pre,twocolumn,superscriptaddress]{revtex4-2}
\usepackage[colorlinks=true,allcolors=blue]{hyperref}
\usepackage{amssymb}
\usepackage{amsmath}
\usepackage{graphicx}
\usepackage{amsfonts}
\usepackage{xcolor}
\usepackage{bm}
\usepackage{hhline}
\usepackage{newtxtext}
\usepackage{graphicx}
\usepackage{xcolor}

\newcommand{\bfE}{\mathbf{E}}
\newcommand{\clE}{\mathcal{E}}
\newcommand{\inn}{\textit{in}~}
\newcommand{\out}{\textit{out}~}

\newcommand{\abs}[1]{\left|#1\right|}
\newcommand{\mean}[1]{\left\langle#1\right\rangle}

\renewcommand{\thetable}{\arabic{table}}

\begin{document}

\title{Nonlinear effects in a strongly coupled Nanoelectromechanical System}

\author{Narges Tarakameh Samani}
\affiliation{Dipartimento di Fisica, Politecnico di Milano, Piazza Leonardo da Vinci 32, I-20133 Milan, Italy}

\author{Farhad Shahbazi}
\affiliation{Department of Physics, Isfahan University of Technology, Isfahan 84156-83111, Iran}

\author{Mehdi Abdi}
\affiliation{Department of Physics, Isfahan University of Technology, Isfahan 84156-83111, Iran}

\begin{abstract}
Controlling nonlinear effects in micro-- and nano-electro-mechanical systems is essential for unlocking their full potential in sensing, signal processing, and frequency control. In this study, we develop a voltage-dependent Hamiltonian framework for a nanoelectromechanical resonator with two strongly coupled vibrational modes, representative of a nanostring platform. The mode frequencies and couplings of the system are tuned electrostatically using a DC voltage, which also controls the strength of the interactions. Our theoretical model reproduces the experimentally observed avoided crossing in the absence of an AC drive and generates tunable frequency-comb spectra when a parametric drive is applied. By scanning the DC voltage, we generate a phase diagram that links comb formation and sharp regime boundaries to underlying bifurcations, multi-stability, and attractor switching. Phase-resolved diagnostics based on a Kuramoto order parameter, together with autocorrelation and Poincaré analyses, quantify coherence and critical slowing down near these transitions. 
We further explore the relationship between nonlinear coupling, parametric excitation, and stability transitions within a single device of experimental relevance and establish a dynamical framework for engineering nanoelectromechanical resonators that offer enhanced tunability, functionality, and a predictive link to experimental outcomes.

\end{abstract}
\maketitle

%
%
\section{Introduction}%

Nonlinearity is a fundamental characteristic of many physical systems, where the relationship between the input and output is not simply proportional, but can vary dramatically with the operating conditions, leading to rich and often unexpected dynamics. Nonlinearity plays a crucial role in various fields, ranging from materials science to quantum systems. ~\cite{ignatenko2019nonlinear,melkikh2015nonlinearity}. 
These effects have mostly been explored in optical setups, where the modes of the optical cavities exhibit high-quality oscillations. Naturally, photons do not interact with one another. Consequently, methods have been developed to enable interactions between them, using materials with nonlinear susceptibilities. Unlike a vacuum, these materials provide an environment where polarizability is dependent on the field strength. This concept was subsequently extended to other oscillators, including mechanical vibrations. ~\cite{huang2020planar,de2016tunable,rega2022nonlinear}.
The idea of engineered nonlinearity has spread into other systems, including Nano-Electro-Mechanical Systems (NEMS)~\cite{craighead2000nanoelectromechanical,ben2023characterisation}. 

NEMS have become indispensable in modern science and technology, thanks to their compact size, low energy consumption, and ability to perform tasks that are unachievable with traditional macro-scale systems. ~\cite{asadi2018nonlinear,dinh2023,rhoads2010nonlinear,ekinci2005nanoelectromechanical}. By integrating sensing, actuation, and signal processing into a unified system, NEMS open up new possibilities in fields such as precise sensing, frequency-selective filtering, inertial navigation (e.g., gyroscopes), and high-density integrated circuits~\cite{wei2021recent,samanta2015nonlinear}. Their compact structure and low mass enable them to operate at high frequencies with remarkable sensitivity, making them ideal for detecting extremely small variations in physical quantities such as force, mass, or pressure on the nanoscale~\cite{Rieger2012,Tu2020}. This capability is essential for applications in scanning probe microscopy, quantum technologies, and nanopositioning systems~\cite{Jia2021,braakman2019force}. Furthermore, their low power requirements make them well-suited for energy-constrained environments, supporting the development of portable and autonomous electronic systems.
To exploit the full potential of such systems, it is essential to understand the fundamental physical principles that dictate their behavior, with nonlinearity playing a central role. This aspect is the basis for many of the unique and complex dynamics observed in these systems. In NEMS, nonlinear effects can arise from several sources, including the intrinsic properties of the materials, geometric nonlinearities due to large deformations, and external forces such as electrostatic interactions~\cite{samanta2018tuning,tang2005physical}. 

Nonlinear effects in NEMS become significant at large oscillation amplitudes, leading to a variety of complex dynamical behaviors. One consequence of this complexity is the possibility of critical transitions in the system dynamics, which can lead to abrupt changes when small variations in conditions push the system past a threshold. This is especially pertinent in NEMS, where maintaining system stability and predictability is crucial for reliable operation~\cite{pavithran2023tipping,bury2020detecting}. These transitions can often be predicted through early warning signals, such as critical slowing down or increased variance of the parameters in the system, which are key indicators for predicting and potentially preventing unwanted dynamical shifts~\cite{Early}. Understanding and managing these nonlinear phenomena is therefore vital not only to avoid instability but also to exploit nonlinear effects in many applications.

One of the significant consequences of nonlinear dynamics is the generation of frequency combs, which are a series of equally spaced spectral lines in the frequency domain ~\cite{mosca, bifurcation, ochs2022frequency}. Integrating frequency comb generation into NEMS platforms allows the development of highly sensitive, compact, and tunable devices~\cite{rahmanian2025nems}. Several studies have shown that the frequency combs generated in NEMS systems are tunable and their spectral properties arise from nonlinear mode interactions and amplitude-dependent responses~\cite{park2019formation,chiout2021multi}.These observations highlight the key role of intermodal coupling and nonlinear resonance conditions in shaping the comb structure, offering a route for dynamically controlling the comb spacing by adjusting system parameters, such as the drive amplitude or frequency~\cite{wu2025limit,ochs2022frequency}.

In this paper, we present a theoretical model to interpret certain experimental observations reported in~\cite{Seitner2017}. The system under study is a free-standing nanoscale beam subject to an electric field, vibrating in two spatially orthogonal modes.
The frequency of these two flexural modes is highly tunable through the external electric field, which also couples them.
We investigate the vibrational dynamics of this system by applying numerical tools to our model.  Specifically, we calculate the phase coherence of the modal vibrations, which reveals the presence of bifurcation in certain parameter configurations. We also demonstrate that the system can generate a tunable frequency comb in the parameter regime where nonlinear effects are prominent.
This tunability is the result of the voltage-dependent frequency disparity between the in-plane and out-of-plane modes. In contrast to previous studies, the comb spacing in our model is tuned via the applied DC voltage, which modulates the coupling between the in-plane and out-of-plane modes~\cite{wu2025limit,ochs2022frequency}. This tuning mechanism offers enhanced precision and lower energy consumption than previously studied methods that rely on altering drive amplitude or frequency. 
To further explore the system's dynamics, we construct Poincaré maps at different DC voltages, revealing the emergence of multiple stability regimes.
Our theoretical findings closely align with the experimental observations and provide a consistent framework for interpreting and utilizing the device.

%
%
\section{Model}
\begin{figure}[t]
\includegraphics[width=1\linewidth]{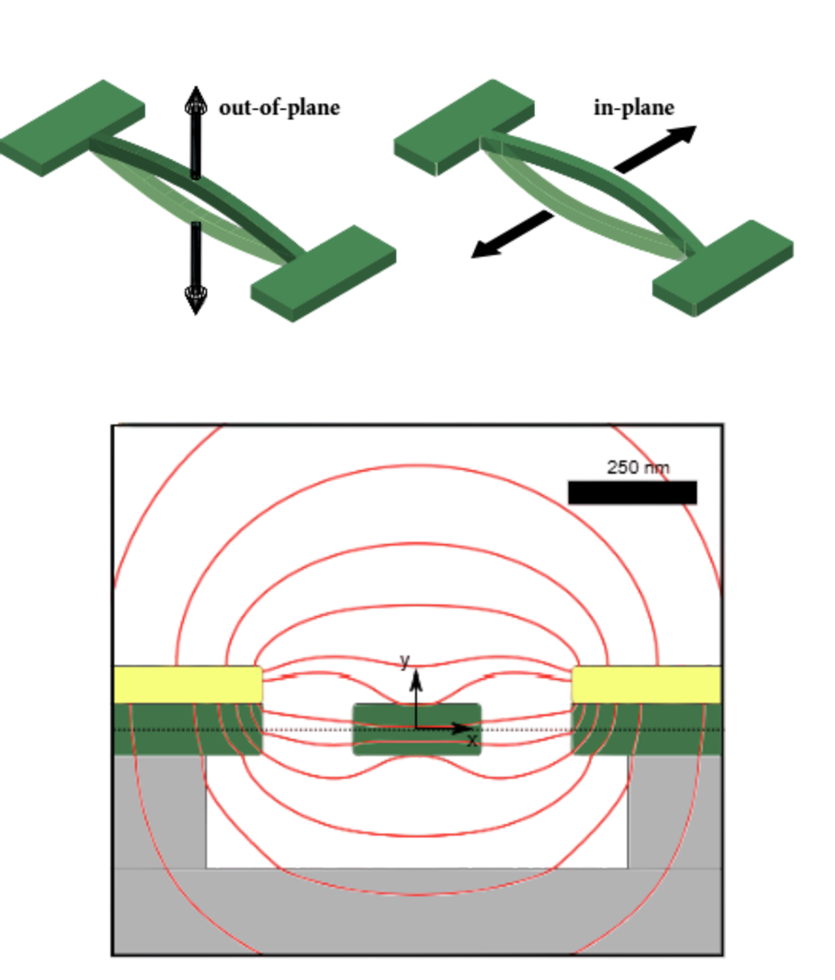}
\caption{%
Top: The nanobeam oscillating in two orthogonal modes. Bottom: Geometry of the nanobeam inside the device in the \textsc{comsol m}ultiphysics\textsuperscript\textregistered~simulations and the simulated electric field lines.
}%
\label{fig:geometry1}
\end{figure}

The theoretical model developed in this study is based on the experimental setup investigated by Faust et. Al~\cite{Faust2012}.The system consists of a doubly clamped silicon nitride nanobeam with a length of $60$~\textmu m  and a cross-sectional area of  $27$~\textmu m$^2$, under a large tensile force applied at both ends.
The nanobeam is free to undergo flexural oscillations in two orthogonal directions: in-plane and out-of-plane. Refer to Fig.~\ref{fig:geometry1} for an illustration.
Due to the high tensile stress imposed by the boundary conditions, the beam behaves like a string, and under ideal conditions, the in-plane and out-of-plane fundamental modes would have identical frequencies. However, slight deviations from this ideal regime, arising from bending tensions, lead to a small difference between the two mode frequencies.
Furthermore, the beam's elastic properties inherently give rise to nonlinear behavior, leading to third-order nonlinear motions at large oscillation amplitudes.


Two gold electrodes are positioned on each side of the beam. These electrodes are used to apply an external electric field to the beam. Experimental results show that the frequencies of the in-plane and out-of-plane modes can be tuned by varying the DC voltage applied to the system. This effect is due to the electrostatic interactions between the external field and the beam. As a result of the motion in this common electrostatic field, the \inn and \out modes are coupled. Observations show that this coupling produces an avoided crossing in the frequency spectrum of the system, consistent with reports in ~\cite{Rieger2012}. To further examine the various behaviors of the nanobeam in this electrostatic configuration, we divided the modelling problem into two parts: The intrinsic mechanical and the external electric effects.



\subsection{Intrinsic mechanical properties}
The dynamics of elastic beams can be well explained by the Euler-Bernoulli equation:
\begin{equation}
\rho S\frac{\partial^2\bm\xi}{\partial t^2}=-EI\frac{\partial^4\bm\xi}{\partial z^4} +(T_0 +\Delta T)\frac{\partial^2\bm\xi}{\partial z^2},
\label{euler}
\end{equation}
where $\rho$ is the material density, $S$ is the cross-sectional area of the beam, $E$ is the elasticity modulus of the material (silicon nitride in our system), $I$ is the second moment of area of the beam cross-section (this parameter is different for the two perpendicular oscillation modes), $T_0$ is the built-in tensile force at the boundary of the beam, and $\Delta T$ is the tensile force resulting from its bending. $\bm\xi(z)$ describes the lateral deflection of the beam at $z$ position in the direction of the beam axis. We decomposed $\bm\xi(z)$ into two orthogonal coordinates $\bm\xi=(x,y)$. Considering the plane of setup as $xz$-plane, we define the \inn plane mode for the $x$-axis vibration and \out of plane mode as $y$-axis vibration, see Fig.~\ref{fig:geometry1}.
In our case, the beam was subjected to a large tensile force:
\begin{equation}
EI\abs{\frac{\partial^4\bm\xi}{\partial z^4}} \ll T_0\abs{\frac{\partial^2\bm\xi}{\partial z^2}}.
\end{equation}
Therefore, as mentioned before, the mechanical resonator behaves like a string, and one expects equal frequency of oscillations for both \inn and \out modes determined by  $T_0$: $\omega_n^2\approx n^2\pi^2T_0/(2mL)$, where $m\approx0.5\rho SL$ is the effective mass, and $L$ is the length of the string. However, bending tension still imposes corrections on the system, and in practice, the two modes have different frequencies.
Because the nanobeam exhibits oscillations with a very high quality factor, its normal modes are well separated from each other and can thus be treated as isolated modes. For our purposes, only the fundamental modes of the two degrees of freedom are driven and contribute to the dynamics of the device.
The Hamiltonian describing the dynamics of the modes and their coupling is:
\begin{equation}
H_0= H_{x}^{(0)} +H_{y}^{(0)} +\lambda_0 x^2y^2.
\label{mechanical}
\end{equation}
Here, we have the following nonlinear Hamiltonian for each mode
\begin{subequations}
\begin{align}
H_{x}^{(0)} = \frac{p_x^2}{2m} +\frac{1}{2}m\omega_{x,0}^2 x^2 +\frac{\beta_0}{4}x^4, \\
H_{y}^{(0)} = \frac{p_y^2}{2m} +\frac{1}{2}m\omega_{y,0}^2 y^2 +\frac{\beta_0}{4}y^4,
\label{duffing}
\end{align}
\end{subequations}
where $\beta_0 \approx \pi^4 ES/(2L^3)$ is the Duffing nonlinearity factor of the mechanical modes and $\lambda_0=\beta_0/2$ is the nonlinear coupling rate between the modes. The intermode coupling can be observed as a consequence of the nonlinear oscillations of the nanobeam. The effect of this interaction has been observed in the experimental results and explained as four-wave mixing processes~\cite{Seitner2017}.

\subsection{External electric configuration}
The electric field produced by the gold electrodes affects the motion of the nanobeam by imposing a position-dependent force. Because silicon nitride is a dielectric, the problem is similar to that of a dielectric slab between two conductors. In general, the problem is not static as the beam oscillates; however, it can be reduced to an electrostatic problem by assuming that the time scale of electric field reformation is much shorter than the period of the mechanical oscillations. This assumption remains valid due to the use of stable voltage sources and high-conductivity electrodes. Additionally, the beam is modeled as a homogeneous dielectric slab.
When a dielectric is inserted into an external electric field under a fixed potential condition, the electrostatic energy difference of the system is given by
$U =\frac{\epsilon -\epsilon_0}{2}\int_V\bfE \cdot\bfE_0dv$,
where $\mathbf{E}_0$ and $\mathbf{E}$ are respectively the electric field before and after insertion of a dielectric object with volume $V$, and the permittivity $\epsilon$.
To simplify the general formula for our specific case, we assume the nanobeam undergoes only pure translational motion—in-plane and out-of-plane—without rotation. This assumption is justified by the tight clamping conditions, which effectively constrain rotational degrees of freedom. Consequently, the integral limits depend solely on the beam’s center-of-mass position, denoted by $\bm\xi$:
\begin{equation}
U(\mathbf{\bm\xi})=\frac{\epsilon -\epsilon_0}{2}\int_{V(\bm\xi)}\bfE(\mathbf{\bm\xi+r'})\cdot\bfE_0(\mathbf{\bm\xi+r'})d^3\mathbf{r'},
\label{electro}
\end{equation}
where $\mathbf{r'}$ is the radial distance of each point of the cross-section from its center of mass and $\bm\xi$ denotes the coordinate of the center of mass in the lab frame. We consider the equilibrium position of the mechanical resonator as the origin of the lab frame $\bm\xi{=}0$.
By assuming small deviations for $\bm\xi$ from the equilibrium position and substituting the Taylor expansion of the electric field around this point we have  $\bfE(x+x',y+y') =\bfE(x',y') +(x+x')\partial_x\bfE(x',y')+(y+y')\partial_y\bfE(x',y') +\cdots$, and similarly for $\mathbf{E_0}(x+x',y+y')$.
By keeping the expansion terms up to the fourth order and incorporating them into Eq.~(\ref{electro}) one obtains:
\begin{equation}
U(x,y) =\sum_{i=0}^{4}\sum_{j=0}^{4}U_{i,j}(\frac{x}{d})^i(\frac{y}{d})^j,
\label{electric}
\end{equation}
where $d$ is the distance between the electrodes and a measure of the system dimension.
The coefficients $U_{i,j}$ depend on the specific geometry of the electrodes and the nanobeam, which are obtained from finite element simulations carried out using \textsc{comsol m}ultiphysics\textsuperscript{\textregistered}.
A simplified 2D configuration resembling the experimental condition is simulated for $d=600$ nm with $-85$ nm out of plane and $-10$ nm in-plane offsets when the electric voltage is set to 14~volts [Fig.~\ref{fig:geometry1}]. The expansion coefficients in Eq.~(\ref{electric}) are then computed, which are listed in Table~\ref{tab:coefs} with reference energy $U_{0,0} = 1.37 \times 10^{-8} V_{\rm DC}^2~[\text{gr}/(\text{nm}^2\cdot \text{\textmu s}^2)$].


\renewcommand{\arraystretch}{1.5}
\begin{table}[tb]
\begin{center}
\begin{tabular}{|c|c|c|c|c|}
\hhline{-|-|~}
$\widetilde U_{1,0}$ & $\widetilde U_{0,1}$ \\
\hhline{-|-|~}
$+0.1937$ & $+0.8598$ \\
\hhline{-|-|-|~}
$\widetilde U_{2,0}$ & $\widetilde U_{1,1}$ & $\widetilde U_{0,2}$ \\
\hhline{-|-|-|~}
$+5.8422$ & $+0.4070$ & $-4.3829$ \\
\hhline{-|-|-|-|~}
$\widetilde U_{3,0}$ & $\widetilde U_{2,1}$ & $\widetilde U_{1,2}$ & $\widetilde U_{0,3}$ \\
\hhline{-|-|-|-|~}
$+1.9371$ & $+12.332$ & $-1.9673$ & $-5.9498$ \\
\hline
$\widetilde U_{4,0}$ & $\widetilde U_{3,1}$ & $\widetilde U_{2,2}$ & $\widetilde U_{1,3}$ & $\widetilde U_{0,4}$ \\
\hline
$+29.335$ & $+7.3650$ & $-59.223$ & $-5.3364$ & $+12.775$ \\
\hline
\end{tabular}
\end{center}
\caption{%
Simulated values for the normalized coefficients in Eq.~(\ref{electric}), i.e., $\widetilde U_{i,j} \equiv U_{i,j}/U_{0,0}$.
}%
\label{tab:coefs}
\end{table}
\renewcommand{\arraystretch}{1.0}
\subsection{The system dynamics}
By integrating the two aspects of the system discussed in previous sections, we can derive its total Hamiltonian.
It should be noted that to drive the system, the electric field includes an AC component alongside the DC component.
Hence, $\bfE(t)=\bfE^{(0)} +\bfE^{(1)}\cos(\Omega t)$ with the drive frequency $\Omega$.
The full Hamiltonian is thus $H =H_{x} +H_{y} +H_{x,{\rm d}} +H_{y,{\rm d}} +H_{\rm int}$. Here, $H_{x} = H_{x}^{(0)} +\sum_{i=2}^4 U_{i,0}(x/d)^i$ and $H_{y}=H_{y}^{(0)} +\sum_{j=2}^4 U_{0,j}(y/d)^j$ are the static Hamiltonian for the \inn ($x$) and \out ($y$) modes.
The drive terms are:
\begin{subequations}
\begin{align}
H_{x,{\rm d}}&=\big[1+\clE\cos(\Omega t)\big]^2\sum_{i=1}^4 \frac{U_{i,0}}{d^i}x^i, \\
H_{y,{\rm d}}&=\big[1+\clE\cos(\Omega t)\big]^2\sum_{j=1}^4 \frac{U_{0,j}}{d^j}y^j,
\end{align}
\end{subequations}
where $\clE =|\bfE^{(1)}|/|\bfE^{(0)}|$ is the ratio of  AC amplitude to the DC part.

The interaction Hamiltonian is composed of the cross-talk terms in various orders
\begin{align}
H_{\rm int} &=g(t)xy +\Lambda(t)x^2y^2 \nonumber\\
&+G_{21}(t) x^2y +G_{12}(t) xy^2 \nonumber\\
&+G_{31}(t) x^3y +G_{13}(t) xy^3,
\label{10}
\end{align}
where the coupling rates are defined as follows:
The most prominent contribution comes from the bilinear coupling
$g(t) =\big[1+\clE\cos(\Omega t)\big]^2\frac{U_{1,1}}{d^2}$,
while the quartic $x^2y^2$ intermode interaction rate is
$\Lambda(t)=\lambda_0+\big[1+\clE\cos(\Omega t)\big]^2\frac{U_{2,2}}{d^4}$
where $\lambda_0$ is intrinsic [see Eq.~(\ref{mechanical})].
There is a similar relation between the other coupling rates and the $U_{i,j}$ coefficients:
$G_{ij} = \big[1+\clE\cos(\Omega t)\big]^2\frac{U_{i,j}}{d^{i+j}}$.
The total Hamiltonian implies that the system dynamics can be controlled in two ways.

First, by changing the DC voltage at the electrodes, the thermally excited modes can be tuned towards an avoided crossing. This was captured by the simulations based on our model. The modified frequency of the \inn and \out modes are given by $\omega_x^2 = \omega_{x,0}^2+2U_{2,0}/(md^2)$ and $\omega_y^2 = \omega_{y,0}^2+2U_{0,2}/(md^2)$, respectively. In addition, nonlinearities can detune the resonance frequency of the modes when driven at large amplitudes.
Second, when the system is driven by a parametric AC electric field.
By exciting the system at the drive frequency $\Omega \approx \omega_x +\omega_y$, the $x$ and $y$ modes experience a set of parametric nonlinear interactions.
This is achievable because none of the higher-order vibrational modes have a frequency close to $\Omega$.
Therefore, this drive only disturbs the two lowest mechanical modes.
The values of the Hamiltonian parameters are listed in Table~\ref{tab:coefs2}.  
For this choice of frequency, the $xy^2$ and $x^2y$ interactions are off-resonance and can be neglected.

\begin{table}[t]
\centering
\renewcommand{\thetable}{\arabic{table}} 
\setcounter{table}{1} 

\begin{tabular}{|c|c|c|c|}
\hline
\textbf{Parameter} & \textbf{Relation} & \textbf{Value} & \textbf{Unit} \\
\hline
$L$ & -- & $60 \times 10^3$ & nm \\
$S$ & -- & $100 \times 250$ & nm$^2$ \\
$Y$ & -- & $160 \times 10^{-9}$ & gr/(\textmu s$^2$·nm) \\
$\Delta Y$ & -- & $1.75235 \times 10^{-9}$ & gr/(\textmu s$^2$·nm) \\
$\rho$ & -- &  $2.8 \times 10^{-21}$ & gr/nm$^3$ \\
$m$ & $\frac{1}{2} \rho S L$ & $2.1 \times 10^{-12}$  & gr \\
$T_0$ & $\Delta Y \cdot S$ & $4.38088\times 10^{-5}$ & (gr·nm)/\textmu s$^2$ \\
$\omega_0$ & $\sqrt{\pi^2 T_0/2 m L}$ & $2 \pi \cdot6.5925$ & \textmu s$^{-1}$ \\
$\lambda_0$ & $\pi^4 Y S/12 L^3$ & $1.50323\times 10^{-16}$ & gr/(nm$^2$·\textmu s$^2$) \\
$\beta_0$ & $2 \lambda_0$ & $3.00645\times 10^{-16}$ & gr/(nm$^2$·\textmu s$^2$) \\
$d$ & -- & 600 & nm \\
$\omega_x$ & $\omega_0 - \Delta$ &  $2 \pi \cdot6.13532$ & \textmu s$^{-1}$ \\
$\omega_y$ & $\omega_0 + \Delta$ & $2 \pi \cdot7.04968$ & \textmu s$^{-1}$ \\
$\Omega$ &--& $2 \pi \cdot 13.208$ & \textmu s$^{-1}$ \\
$Q$ & -- & $10^{3}$ & – \\
$\gamma$ & $\omega_0/Q$ & $4.14219\times 10^{-2}$ & \textmu s$^{-1}$ \\
\hline
\end{tabular}
\caption{List of parameters used in the model. $\Delta$ is introduced as a free parameter, inferred from experimental agreement.}
\label{tab:coefs2}
\end{table}


The equations of motion are derived from the Hamilton equations.
The full dynamics of the system are obtained by adding thermal noise and damping.
This, in turn, is accounted for through the Langevin equations of motion:
\begin{subequations}
\begin{align}
&\ddot{x}+\gamma\dot{x}+\frac{1}{m}\frac{\partial H}{\partial x}=\eta_x, \\
&\ddot{y}+\gamma\dot{y}+\frac{1}{m}\frac{\partial H}{\partial y}=\eta_y,
\end{align}%
\label{langevin}%
\end{subequations}%
where $\gamma \approx \frac{\omega_{x,0}}{Q} \approx \frac{\omega_{y,0}}{Q}$ is the damping rate of the mechanical resonator with a quality factor of $Q$ (the explicit form of equations is too cumbersome and thus is not presented here). Note that since the difference between $\omega_{x,0}$ and $\omega_{y,0}$ is minimal and the system operates in a high-Q regime, we assume that the difference in quality factors for the two modes is negligible compared to the magnitude of Q. Therefore, we model the dissipation with an effectively equal damping rate $\gamma$ and an equal quality factor $Q$ for both modes, while retaining the small frequency mismatch in the Hamiltonian contributions $H_x$ and $H_y$. The noise force $\eta_k$ with $k=x,y$ has a zero-mean Gaussian probability distribution, $\mean{\eta_k}=0$, with correlation function
\begin{equation}
\mean{\eta_k(t)\eta_{k'}(t')}=\frac{2\gamma}{m}k_{\rm B}T \delta_{kk'}\delta(t-t'),
\label{xicorr}
\end{equation}
where, $k_{\rm B}$ is the Boltzmann constant and $T$ is the ambient temperature. Here, $\langle \cdot\rangle$ indicates the time average.
Considering the negligible impact of noise and for the sake of simplicity, the following section examines a noiseless system.

%
%
\section{Results}
This section presents the results obtained through the numerical analysis of Eqs.~(\ref{langevin}) from various perspectives.  
\subsection{Spectrum}
\begin{figure}[t]
\includegraphics[width=1\linewidth]{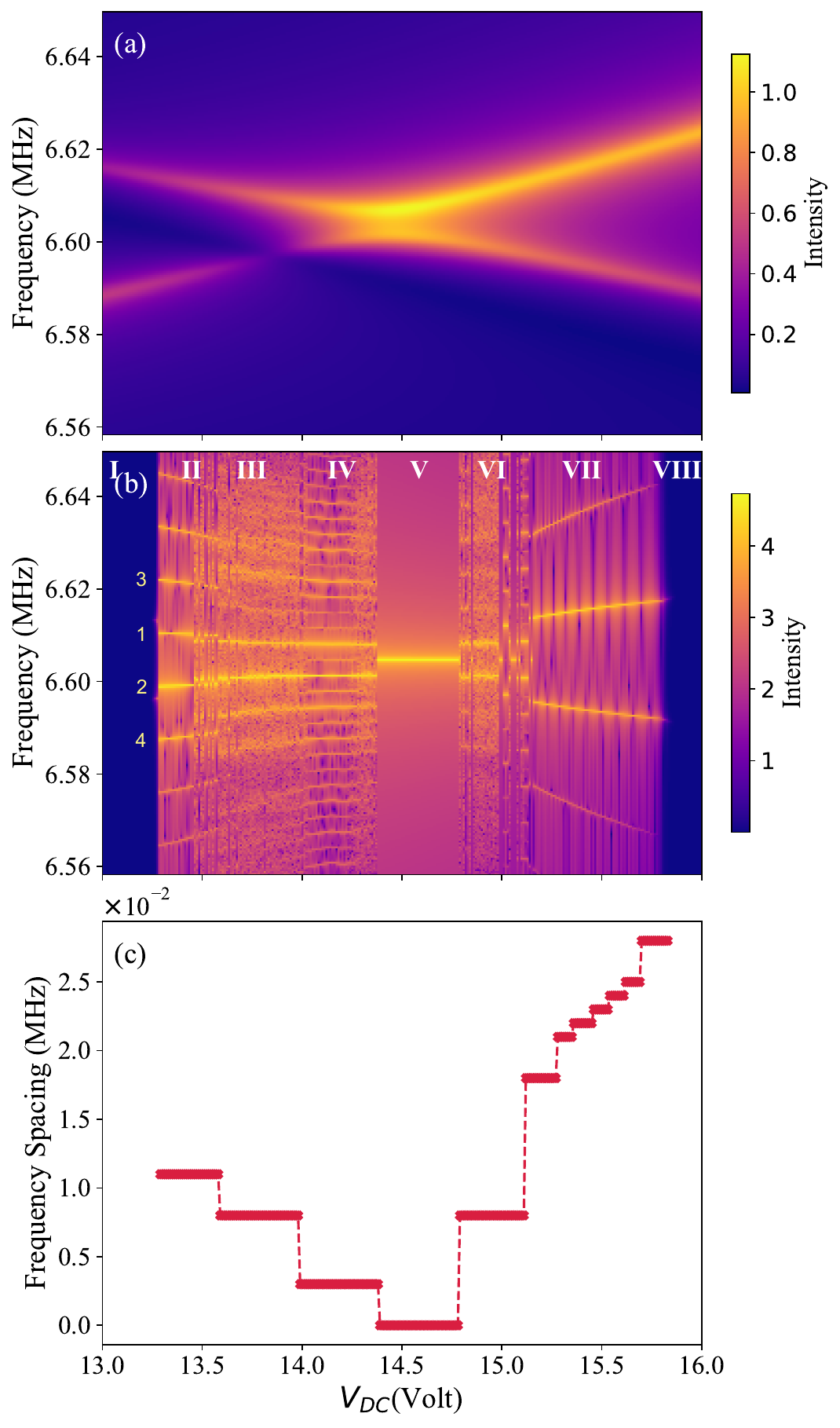}
\caption{%
Frequency spectrum versus $V_{\rm DC}$: (a) for non-driven, and (b) driven cases.
(c) shows the frequency spacing of the combs formed in the driven case.
}%
\label{fig:spec2}
\end{figure}

Starting from the analysis of the static spectrum of the normal modes obtained from the total Hamiltonian, we observe an avoided crossing as $V_{\rm DC}$ is varied in the absence of an external drive ($\clE = 0$), indicating a strong coupling between the two modes. This also confirms that the mode frequencies can be effectively tuned by applying an electrostatic voltage to the electrodes. The corresponding results are shown in Fig.~\ref{fig:spec2}(a).
Applying a parametric drive to the system significantly exposes its nonlinearity, resulting in a more complex spectrum.
Fig.~\ref{fig:spec2}(b) presents the frequency spectra of the signal, which is selected as the sum of two modes of the driven system ($x+y$), plotted against the DC voltage. The drive frequency is set at $\Omega=\omega_x +\omega_y$ with a drive power of 14 dBm.

Around $V_{\rm DC} \lesssim 13$~V the system oscillates very weakly. However, increasing $V_{\rm DC}$ to a value higher than 13.26 V brings the system into a regime where oscillations at several distinct frequencies become possible. As $V_{\rm DC}$ further increases, and as the system approaches the avoided crossing point, it exhibits different behaviors with relatively sharp boundaries and different numbers of resonance frequency lines with varying spacings. This behavior can be linked to the effectively enhanced coupling and subsequent parametric nonlinear processes occurring in the system.

In other words, the parametric drive activates nonlinear interactions, the strength of which is inversely proportional to the frequency spacing.
Hence, as the system approaches the avoided crossing point, these effects become more prominent.
These nonlinear effects, which act on top of the voltage-dependent coupling, lead to complex energy exchange between the modes and result in the appearance of multiple spectral regimes with distinct frequency patterns, as shown in Fig. ~\ref{fig:spec2}(b).

\begin{figure*}[t]
\includegraphics[width=1\linewidth]{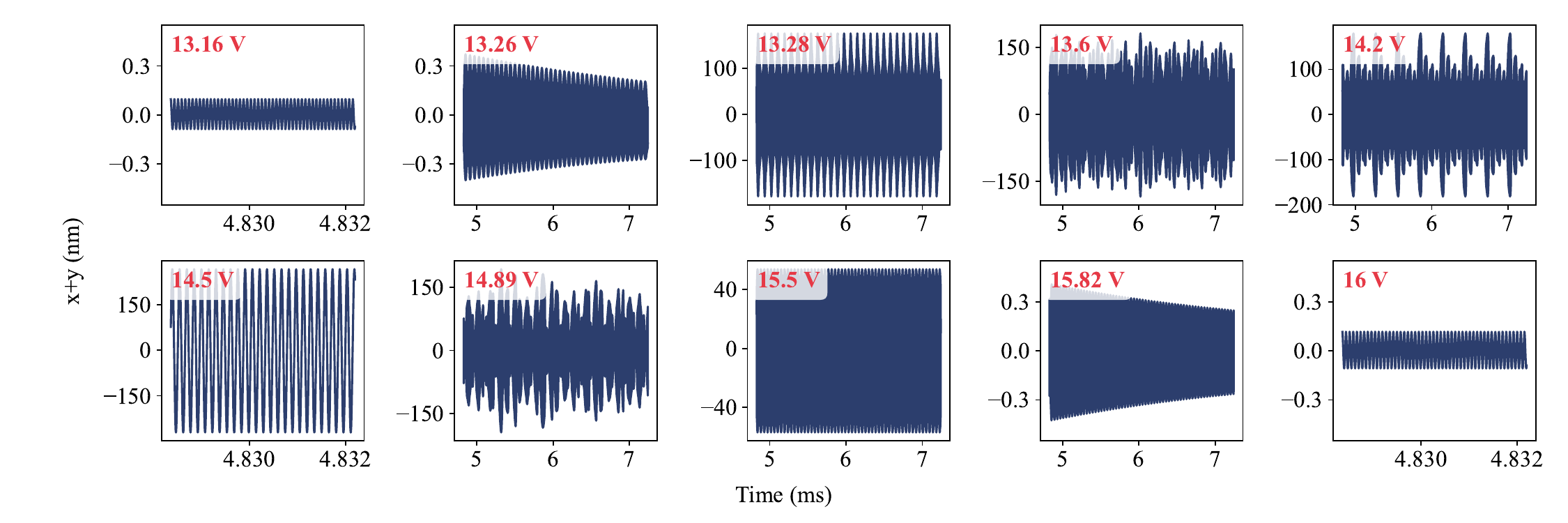}
\caption{%
Variation of the combination of the two mode's position in time, in different $V_{\rm DC}$. 
}%
\label{fig:xyt14}
\end{figure*}

This behavior can be interpreted using the concept of parametric normal mode splitting (PNMS)~\cite{EvaLast}, which provides a physical explanation for the emergence of multiple frequency components in the spectrum.
PNMS explains the energy transfer between the two modes in the system and how an external parameter can influence its behavior. Specifically, because of this phenomenon, the interaction between the two modes causes each frequency line to split into several lines.
In systems with strong light-matter interactions, such as atom-cavity systems, an increase in the coupling strength can lead to a more complex spectral response, characterized by a greater number of observable frequencies. Specifically, when the coupling between atoms and cavity modes is sufficiently strong, a phenomenon known as multi-normal mode splitting can occur. In this regime, multiple cavity modes interact with the atoms, causing each mode to split into several peaks, thereby significantly increasing the number of resonant frequencies in the spectrum of the system.

Similarly, in our system, although the number of frequencies observed in PNMS is typically two for each mode, additional frequencies can emerge due to the strong coupling between the two modes.  In other words, the spectral response with multiple frequencies, as shown in Fig. ~\ref{fig:spec2}(b), emerges from the nonlinear interaction of the modes of the system.
Nonlinear effects become more prominent as the coupling strength increases at strong parametric drive amplitudes, as shown in Eq. ~(\ref{10}).
These nonlinear couplings involve higher-order interaction, leading to the creation of higher harmonics of the original frequencies.
Moreover, they generate new combination frequencies such as $f_1 + f_2$ and $2f_1 - f_2$, which, in turn, add even more complexity to the system's spectrum~\cite{chaosbook}. A very strong coupling may lead to quasi-periodic or chaotic oscillations, introducing broad spectral features or densely packed frequencies~\cite{{shelton2011strong,kashinath2014nonlinear}}. Interestingly, we identified that the relation between the four frequencies shown in Fig.~\ref{fig:spec2}(b) as follows: $f_3=2f_1-f_2$, and $f_4=2f_2-f_1$, which is in agreement with the experimental results presented in ~\cite{{Seitner2017}}.

Within each regime, the frequency components are evenly spaced, indicating the presence of a frequency comb in the system. As quantified in Fig.~\ref{fig:spec2}(c)  the frequency spacing remains constant within a given regime but varies across different regimes. This observation highlights that by adjusting $V_{\rm DC}$, we can effectively tune the generated frequency combs.
\subsection{Order parameter}
The presence of different regimes in the frequency spectra of the system is expected to reflect interesting behaviors in its physical properties, such as the existence of multiple attractors caused by oscillations and vibrational dynamics. The diagrams in Fig.~\ref{fig:xyt14} show the sum of in-plane and out-of-plane displacements of the center of mass (x+y), as derived from Eqs.~(\ref{langevin}), plotted against time for various $V_{\rm DC}$ values. It is evident that varying $V_{\rm DC}$ results in a significant change in oscillation behavior. Studying changes in both amplitude and phase is particularly challenging, because many factors can affect amplitude variations and complicate the analysis. Furthermore, the time series do not exhibit significant changes and are inadequate for determining the transition condition that leads to the sharp boundaries observed in the frequency spectra shown in Fig.~\ref{fig:spec2}(b). Therefore, to determine the stable states (attractors) of the system, it is important to define a parameter that reflects its stability and quantifies the coherence between the two original modes without being affected by amplitude variations.
\begin{figure}[b]
\includegraphics[width=1\linewidth]{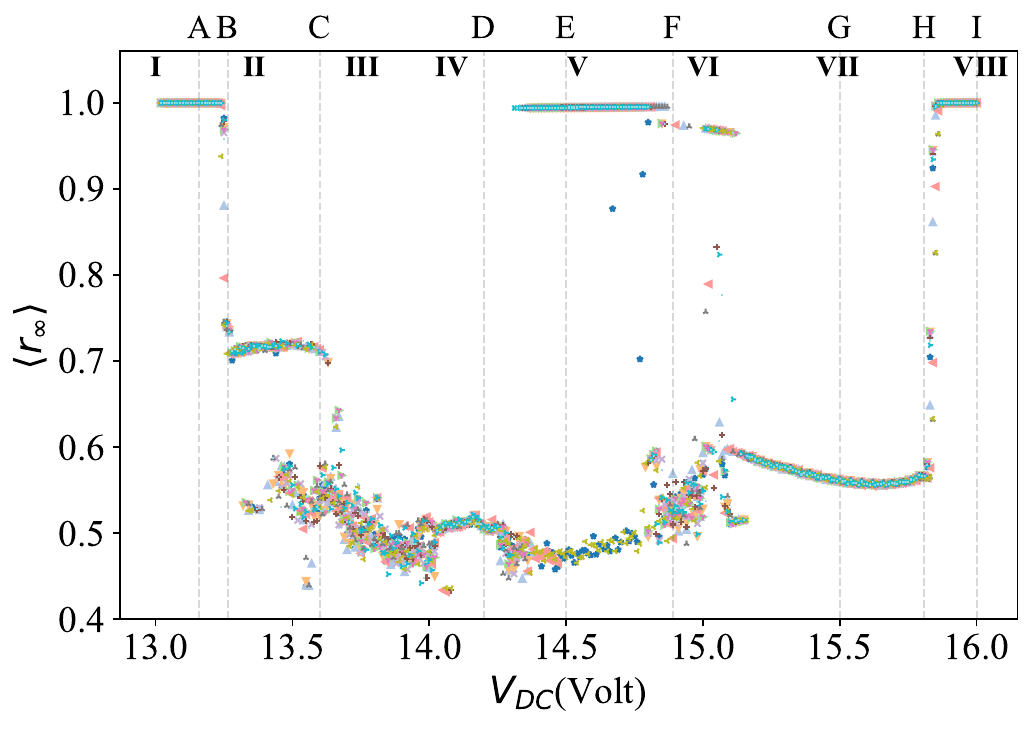}
\caption{%
Variation of the order parameter (considering different initial conditions) shows changes in relative phase in simultaneous voltages, as in the frequency spectrum. The labels 'A' to 'I' mark some key voltage values discussed in this paper: A: 13.16 V, B: 13.264 V, C: 13.6 V, D: 14.2 V, E: 14.5 V, F: 14.89 V, G: 15.5 V, H: 15.808 V, I: 16 V.
}%
\label{fig:r-vdc}
\end{figure}
\begin{figure*}[tbh!]
\includegraphics[width=1\linewidth]{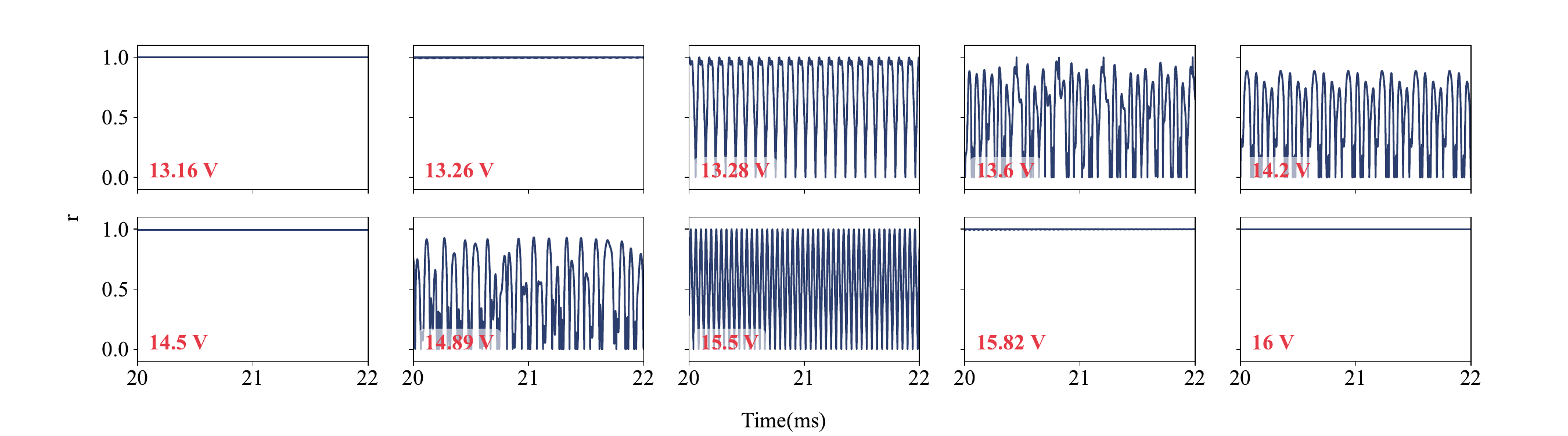}
\caption{%
Evolution of order parameter in time at different $V_{\rm DC}$ s, corresponding to a fixed initial condition, showing transition to new states.  
}%
\label{fig:dynamics}
\end{figure*}

The most significant parameter that contains coherency-related information is the phase difference of the oscillators. Using the signals discussed earlier (x, y), the phase of the oscillators can be obtained using the Hilbert transform~\cite{hilbert}. This approach represents the signal in its analytic form, which is a method of signal processing that separates the amplitude and phase information. The analytic signal $z(t)$ is given by:
\begin{equation}
z(t) = x(t)+i\mathcal{H}[x(t)].
\end{equation}
where $\mathcal{H}[x(t)]$ is the Hilbert transform of $x(t)$. Consequently, using this analytic signal, the phase can be obtained as follows:
\begin{equation}
\theta(t)=\tan^{-1}(\frac{\text{Im}(z(t))}{\text{Re}(z(t))}).
\end{equation}

To quantify the coherence and phase relationship, we use the Kuramoto order parameter, which provides insight into the collective behavior of the system~\cite{mahdavi2025synchronization}. This parameter indicates how well the phases are aligned and whether the system is in an ordered or disordered state.
\begin{equation}
r(t)=\frac{1}{2} \abs{\sum_{j=x,y}e^{i\theta_j(t)}}=\abs{\cos{\frac{\phi(t)}{2}}},
\end{equation}
where $\phi(t)=\theta_x(t) -\theta_y(t)$ is the relative phase between the two modes.
 
In cases where the two oscillators are aligned in phase $(\phi=0)$, the order parameter achieves a value of one. Conversely, when they are in opposite phases $(\phi=\pi)$, the order parameter drops to zero. Fig.~\ref{fig:r-vdc} shows the variation of $\langle r_\infty \rangle$, which is the average of the order parameter in steady state, versus $V_{\rm DC}$ under different initial conditions. These diagrams demonstrate that sharp changes also appear in the system’s order parameter, occurring at the same $V_{\rm DC}$ values as abrupt shifts in frequency spectra. This indicates that changes in the frequency and relative phase occur simultaneously, potentially indicating fundamental transitions, such as bifurcation, within the system. Fig.~\ref{fig:dynamics} shows the time evolution of the order parameter for selected values of \( V_{\rm DC} \), representing different dynamical regimes. The system exhibits clear differences in behavior across these regimes: in some cases, it remains synchronized,; in others, it loses synchronization and becomes non-periodic, and at certain voltages, it returns to periodic behavior. Overall, this figure confirms that the sharp changes observed in the order parameter \( r_{\infty} \) correspond to significant transitions in the system’s underlying dynamics.

To analyze how the system’s dynamics change, we examine the autocorrelation of its phase fluctuations. This approach helps us assess the temporal coherence of the oscillators by calculating the phase autocorrelation function:
\begin{equation}
C(\tau) = \left\langle e^{i[\phi(t) - \phi(t+\tau)]} \right\rangle.
\label{auto}
\end{equation}
and focus on its modulus $A(\tau) = \left| C(\tau) \right|$, which quantifies the persistence of phase coherence over time.
\begin{figure*}[tbh!]
\includegraphics[width=1\linewidth]{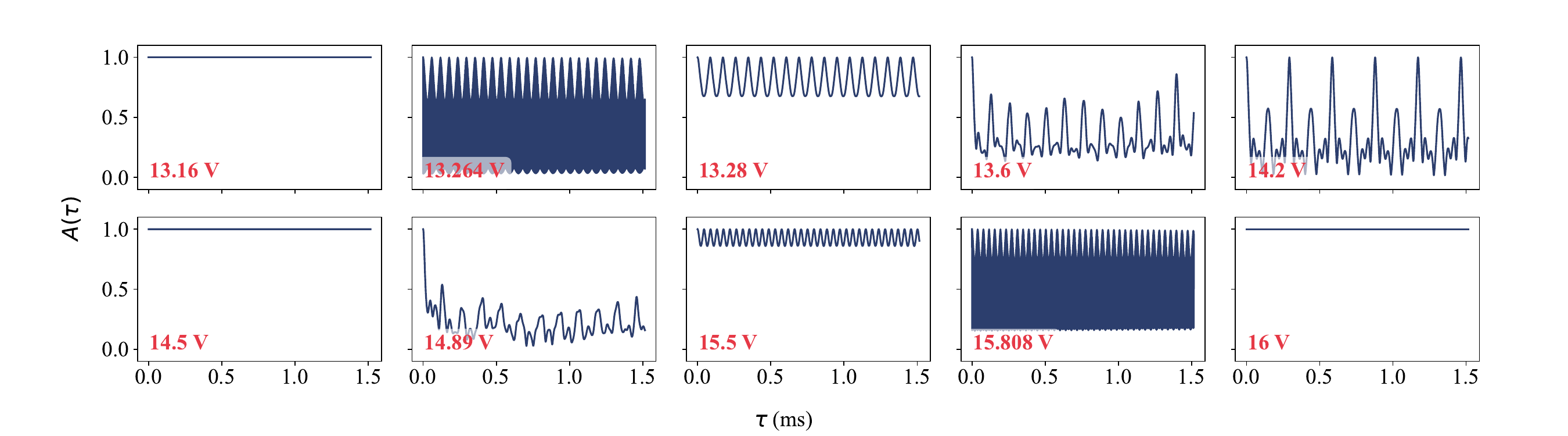}
\caption{%
Autocorrelation of the system obtained by Eq.\eqref{auto}, shown for different $V_{\rm DC}$s corresponding to different regimes.
}%
\label{fig:autocorr}
\end{figure*}
As shown in Fig.~\ref{fig:autocorr}, the behavior of $A(\tau)$ reveals different dynamical regimes. At $V_{\mathrm{DC}} = 13.16\,\mathrm{V},14.5 \,\mathrm{V}, 16\,\mathrm{V}$, $A(\tau)$ remains flat and close to 1, indicating a fully synchronized regime with no significant phase fluctuations. However, at specific voltages, most clearly at \( V_{\mathrm{DC}} = 13.264\,\mathrm{V} \) and \( 15.808\,\mathrm{V} \), we observe the emergence of rapid oscillations in the autocorrelation. This behavior strongly marks a transition zone in the system. These modulations in \( A(\tau) \) reflect the onset of internal frequencies and a breakdown of the synchrony, indicating that the system is entering a qualitatively different dynamical regime, which is likely a quasi-periodic or toroidal state. The abrupt change in the temporal correlation structure can be a signature of an underlying bifurcation~\cite{Early}. 
At $V_{\mathrm{DC}} = 13.28\,\mathrm{V},14.2\,\mathrm{V}$, $15.5\,\mathrm{V}$ the oscillatory character of $A(\tau)$ persists almost without decay, indicating a sustained periodic state. Instead, what can be observed at $V_{\mathrm{DC}} = 13.6\,\mathrm{V},14.89\,\mathrm{V}$ is a relatively fast decay of autocorrelation, which shows a chaotic behavior. All autocorrelation curves are computed from the steady-state time series to ensure that the observed features reflect the long-term system behavior rather than transient phenomena. We now turn our attention to the geometry of the attractors. By analysing the Poincaré sections, we can directly observe how the structure of the system's dynamics changes across these transitions.
\begin{figure*}[tbh!]
\hspace*{-0.9cm} 
\includegraphics[width=1\linewidth]{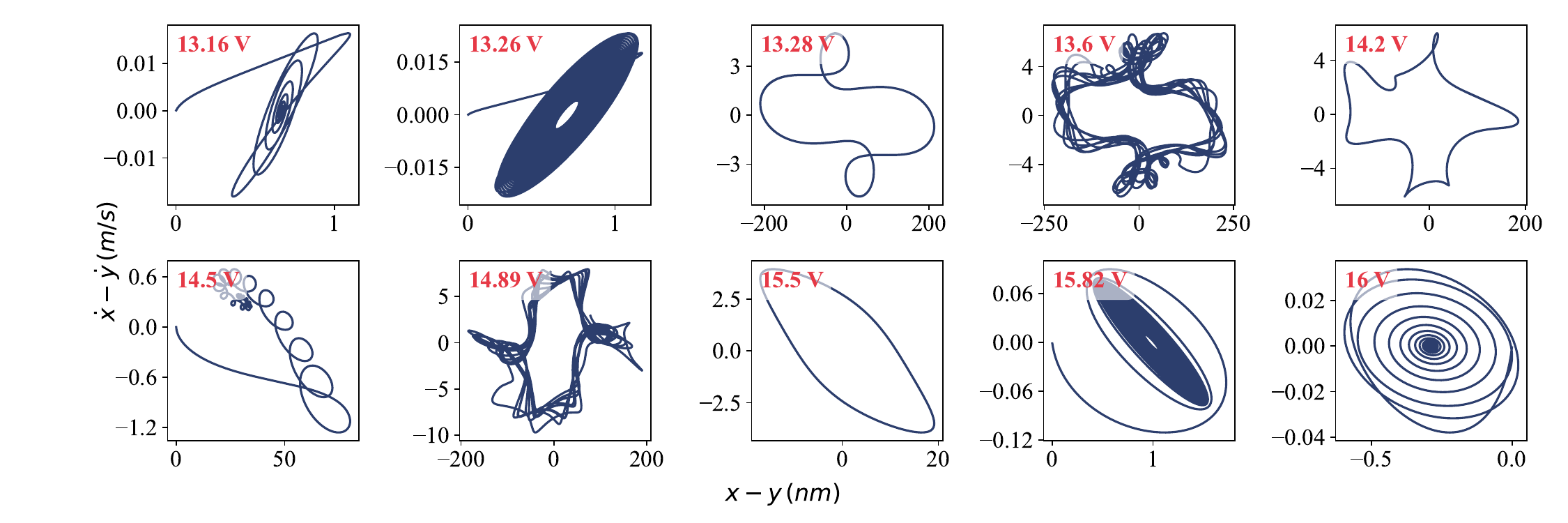}
\caption{%
Poincaré sections showing the intersections of the system's trajectory with the $(x-y, \dot{x}-\dot{y})$ plane.  Each panel corresponds to a different value of $V_{\rm DC}$, representing distinct dynamical regimes of the system.
}%
\label{fig:poincare}
\end{figure*}

\subsection{Poincaré section}

The system's phase space holds important information about the trajectories of the two original modes. Therefore, direct examination of the spatial variation of these modes can help in understanding these fundamental changes. However, as shown in Fig.~\ref{fig:xyt14}, this method is very complicated and not the most effective way to study the system's behavior.

Consequently, we decided to analyze the Poincaré map of the system at the $(x-y,\dot{x}-\dot{y})$ section, which represents the intersection of the trajectory in the phase space with the $(x-y,\dot{x}-\dot{y})$ plane at half of the drive period. This map implies a connection between position and velocity, highlighting the different stable orbits that serve as attractors in the system’s oscillatory behavior and the possible transitions between them. Fig.~\ref{fig:poincare} illustrates the Poincaré sections at selected voltages corresponding to the different regimes shown in both Fig.~\ref{fig:spec2} and Fig.~\ref{fig:r-vdc}, indicating the system's attractors at the respective $V_{\rm DC}$s. For voltages of 13.16\,V, 13.26\,V, 14.5\,V, 15.82\,V, and 16\,V, the Poincaré sections are computed over the entire temporal evolution of the system, covering both the initial transient and the steady-state behavior.


In regime $I$, the attractors display a fixed point in the stationary state of the system, indicating synchronization. Analysis of Fig.~\ref{fig:spec2} and Fig.~\ref{fig:r-vdc} shows that increasing $V_{\rm DC}$ drives the system into a state with a higher frequency count and larger fluctuations in the order parameter, the reason for which can be studied in the Poincaré map. In regime $II$, the system stabilizes in a completely different pattern at considerably higher velocities than in regime $I$, as shown in Fig. ~\ref{fig:poincare}. These limit cycles or chaotic attractors remain mostly unchanged up to voltages in regime $IV$, following similar yet slightly varied trajectories with different repetition times. This creates various frequency components in the spectra. As the system reaches regime $V$, the order parameter jumps to values close to one. Because of this large change, a clear difference is expected in the Poincaré map in this regime. The system's attractor in regime $V$, $V_{\rm DC}$=14.5 V , as observable in Fig. \ref{fig:poincare}, is a fixed point in the stationary state of the system. this regime corresponds to the maximum linear coupling experienced by the system, which is in line with the avoided crossing phenomenon. As $V_{\rm DC}$ increases and the system enters regimes $VI$, $VII$, and $VIII$, the Poincaré map shows noticeable changes in the pattern, which can be used to explain the sudden jumps observed in the frequency spectra. This consistency between the geometric features of the attractors and the temporal behavior of the order parameter, demonstrated in Fig.\ref{fig:dynamics}, supports our identification of the system’s distinct dynamical states. By organizing the system's behavior into distinct regimes, each associated with a distinct attractor, we gain a better understanding of the mechanism of frequency splitting. This highlights the influence of nonlinear dynamics on the system’s response. Overall, this insight allows for more effective control over the frequency-splitting process and the development of more accurate frequency combs.

\subsection{Multi-stability and Bifurcation Signatures}

To pinpoint the nature of the previously discussed transitions, and to check if they involve bifurcations, we analyze additional dynamical signatures typical of nonlinear systems. These include hysteresis, multi-stability, and critical slowing down, all of which can provide strong complementary evidence for the bifurcation behavior as the control parameter \( V_{\mathrm{DC}} \) is varied.
\begin{figure}[tbh!]
\includegraphics[width=1\linewidth]{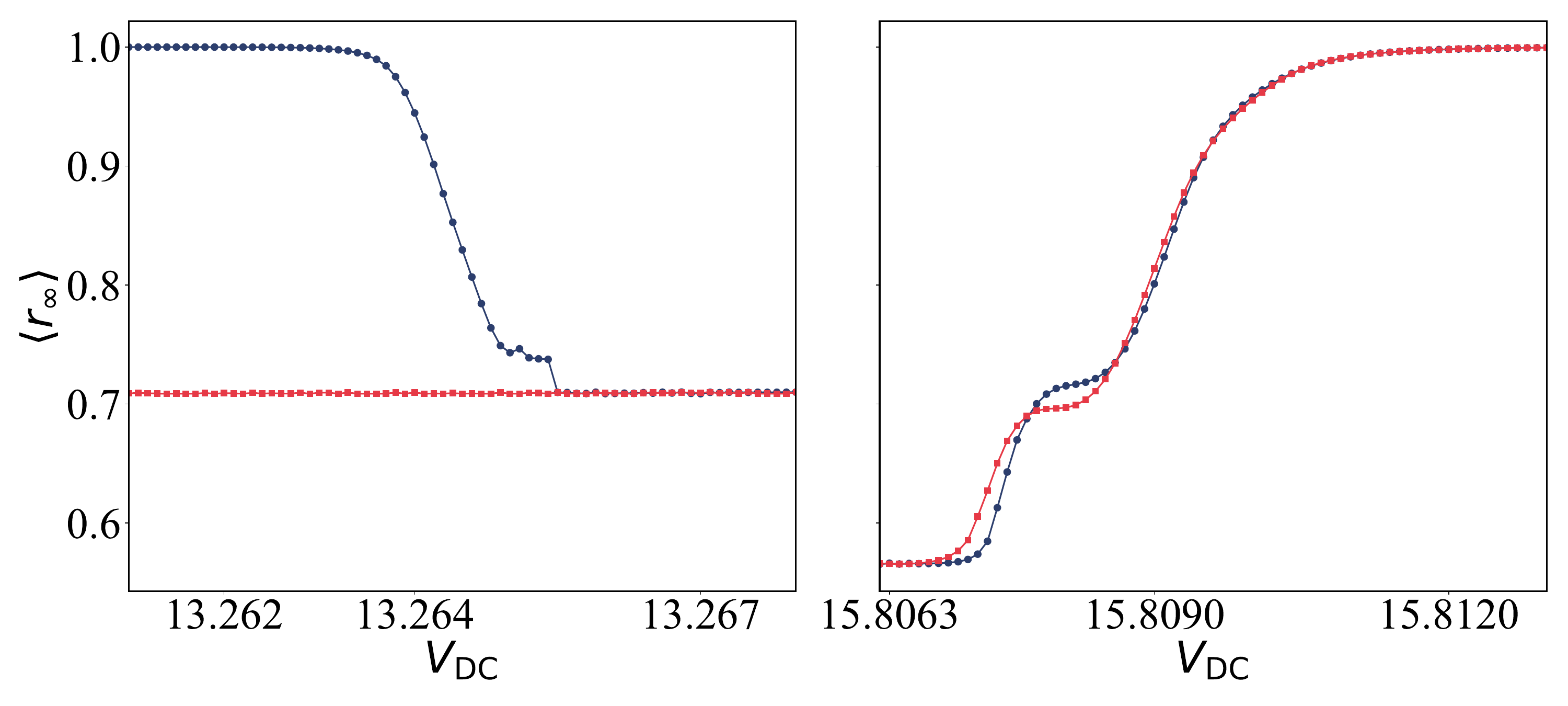}
\caption{%
Hysteresis in the order parameter as a function of the applied DC voltage \( V_{\mathrm{DC}} \). The system exhibits a clear hysteretic loop, indicating multi-stability in certain voltage ranges.‌ Blue color indicates forward and red, backward sweeping.
}%
\label{fig:hysteresis}
\end{figure}
We begin by analyzing how the order parameter responds to slow voltage sweeps in both forward and backward directions. As shown in Fig.~\ref{fig:hysteresis}, clear hysteresis loops appear around two voltage regions: \( V_{\mathrm{DC}} \approx 13.264\,\mathrm{V} \) and \( 15.808\,\mathrm{V} \). In these intervals, the system follows different trajectories depending on the sweep direction, indicating memory effects. This behavior reflects multi-stability, characterized by the coexistence of more than one attractor, and is a well-known signature of saddle-node bifurcation in nonlinear systems~\cite{koike2024bifurcation}. These results align with the previously observed abrupt changes in the autocorrelation function and Poincaré sections, reinforcing the interpretation that the system undergoes bifurcations at these voltages.v
\begin{figure}[h!]
\includegraphics[width=\columnwidth]{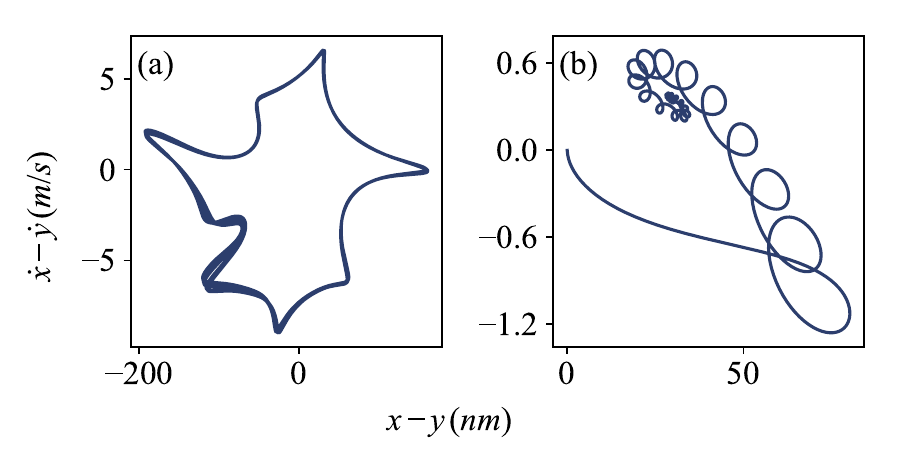}
\caption{%
Poincaré section in regime $V$ with (a) $x_0=1$ nm as initial condition leads to the lower order parameter (shown only in steady state of the system), and (b) $x_0=0$ nm as initial condition leads to a different map (shown by considering transient behavior which end up to a fixed point in the stationary state).
}%
\label{fig:po1}
\end{figure}

Another form of multi-stability appears in regime $V$. As illustrated in Fig.~\ref{fig:r-vdc}, this regime exhibits a high order parameter under certain initial conditions. However, small changes in the initial state result in a sharp drop in the order parameter, causing the system to behave as it does in other regimes. To investigate this, we compute the Poincaré sections for two representative initial conditions. Fig. ~\ref{fig:po1}(a) shows the attractor for \( x_0 = 1\,\mathrm{nm} \), while Fig.~\ref{fig:po1}(b) shows a point attractor for \( x_0 = 0\,\mathrm{nm} \). This confirms the presence of multi-stability in regime~V. Importantly, this case differs from the earlier bifurcations in that the transition between attractors is much more abrupt; even a slight change in the initial condition is sufficient to shift the system into a fundamentally different state. This sharp boundary between the attractors suggests a more sudden or structurally distinct type of bifurcation than those observed near 13.264 V and 15.808 V.
\begin{figure}[tbh!]
\includegraphics[width=\columnwidth]{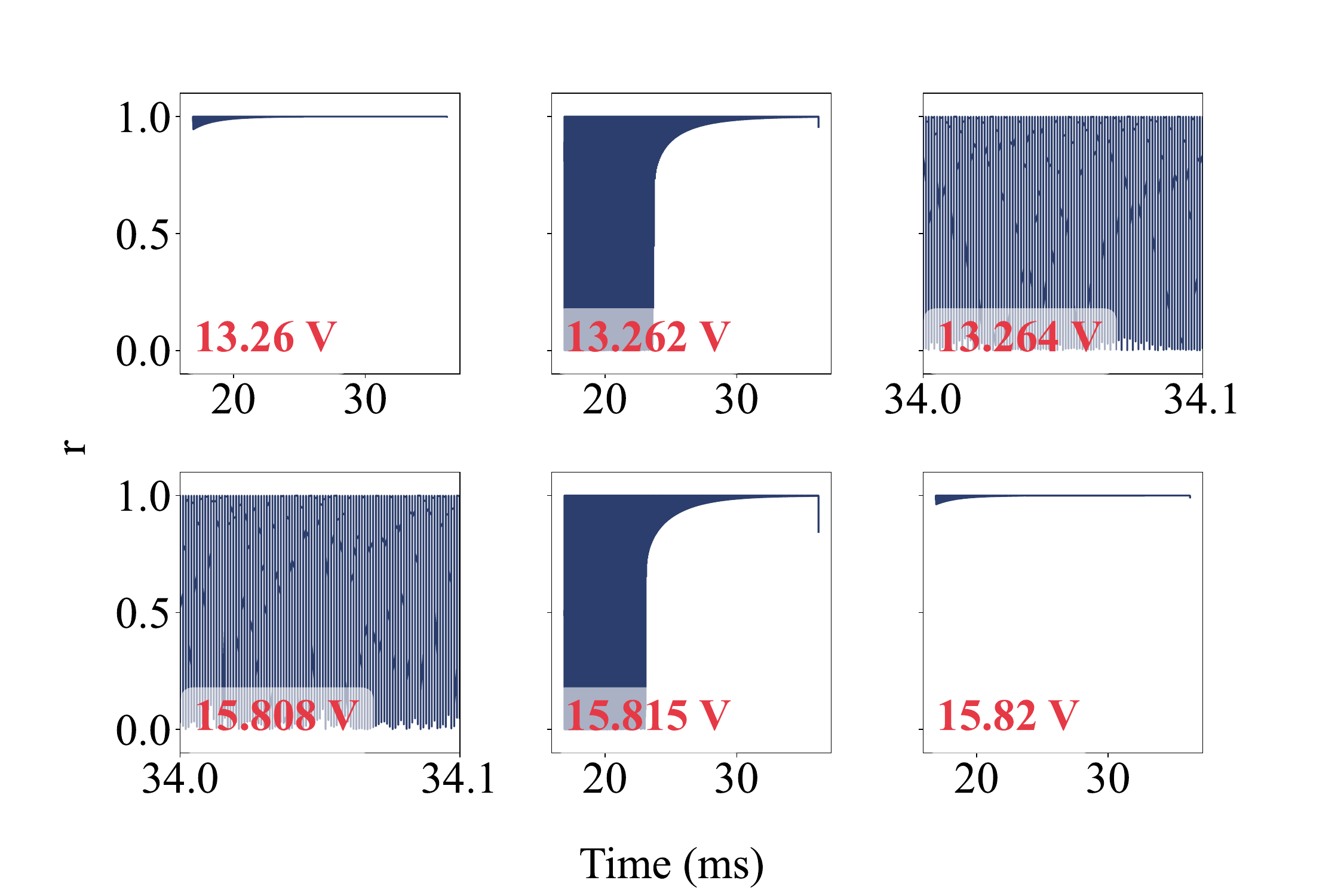}
\caption{%
Dynamics of order parameter in voltages around bifurcation points, showing slowing down by varying voltage}%
\label{fig:slow}
\end{figure}

Finally, we examine the temporal evolution of the order parameter near the earlier bifurcation points. As shown in Fig.~\ref{fig:slow}, the system slows down before settling into a steady state as the voltage approaches \( V_{\mathrm{DC}} = 13.264\,\mathrm{V} \) or \( 15.808\,\mathrm{V} \). In some cases, the system does not fully stabilize at all. This slowing down is consistent with theoretical predictions near bifurcation points, where the recovery time diverges as the system approaches a critical transition~\cite{Early}. Considering these results, the presence of hysteresis, multi-stability, and slowing down provides robust evidence that the system undergoes bifurcations at specific voltages. Therefore, the observed slowing
down before the bifurcation points may be indicator of near-critical behavior in the system.\\

Nonlinear signatures such as bifurcation and multi-stability, as identified in this study, can be probed directly in experiments. For example, in the experiment of Allemeier \emph{et al.}~\cite{Allemeier25} on a high-tension nanostring with two orthogonal flexural modes, the system is driven into the nonlinear regime and its dynamics are mapped by sweeping the drive frequency in the presence of added noise. The noise induces switching between coexisting states, providing a concrete and quantitative protocol to measure multi-stability. While their analysis characterizes multi-stability using probability maps, the same time-series data can also be examined for signatures of critical slowing down—specifically by extracting relaxation and correlation times as the detuning approaches the bifurcation point. A similar protocol can be implemented for the class of devices considered here, offering a direct route to experimentally quantify the nonlinear effects predicted by our model. In addition, the nanostring experiments of Faust \emph{et al}. and Seitner \emph{et al}.~\cite{Seitner2017,Faust2012} use an electrical scheme where mechanical motion is read through electrodes that provide static tuning and rf drive: the displacement of the string modulates the capacitance to an inductively coupled microwave cavity. The resulting change in the cavity transmission is demodulated to yield sharp spectral lines at the mechanical resonance frequencies as functions of drive frequency and dc bias. In such a scheme, the nonlinear bifurcation and multi-stability predicted by our model would manifest as coexisting spectral response branches and abrupt switching between them when the control parameters are swept. Moreover, the demodulated microwave signal can, in principle, be acquired in the time domain and analyzed beyond its average spectrum, allowing one to extract slow switching dynamics and relaxation times of the collective motion, and thereby access the same type of critical slowing down discussed in our theoretical treatment.
\\

\section{Summary and Conclusion}
Through both theoretical and numerical analyses, we investigated the effect of parametric driving on a nanomechanical system with nonlinearly coupled modes. Our work establishes a framework for understanding the complex dynamics of the system. By tuning the frequency difference between the in-plane and out-of-plane modes, controlled through electrostatic interactions under an applied DC voltage, we observed a rich nonlinear response when the AC drive introduced parametric excitation. The system exhibited mode splitting and revealed multiple frequency lines near the avoided crossing point, with frequency combs emerging over specific voltage ranges. These combs show varying line spacings depending on the applied DC voltage, demonstrating the system’s tunability. Phase coherence analysis using the Kuramoto order parameter further highlights the abrupt transitions between stable states, suggesting the presence of bifurcations. We also identified multi-stability in one of the regimes, supported by clear variations in the Poincaré sections under small changes in initial conditions. Moreover, we observed hysteresis, slowing down, and attractor switching behaviors consistent with classical bifurcation phenomena in nonlinear systems. Importantly, previous studies have shown that multiplicative noise can significantly affect such transitions, particularly near bifurcation points, by altering thresholds or stability basins~\cite{pisarchik2014control,phillips2025stabilizing}. Therefore, exploring the role of noise in this system and its influence on bifurcation behavior or transition delays presents a promising avenue for future research.

In conclusion, this study offers insights into strongly coupled nanomechanical systems, revealing nonlinear phenomena such as mode coupling, tunable frequency comb generation, multi-stability, and bifurcation. The complexity and tunability of the system indicate potential applications in precision frequency control and signal generation, while also laying the foundation for future studies of stochastic effects and critical transition in nanoscale dynamics.


%
\begin{acknowledgments}
The authors thank E. M. Weig for helpful comments and discussions and for providing us with the experimental data.
NTS thanks A. Ardeshiri for his support during this work. 

Accepted for publication in Physical Review E (APS).”
DOI: https://doi.org/10.1103/59mg-ffhg
\end{acknowledgments}

\bibliographystyle{apsrev4-2}

\begin{thebibliography}{41}%
\makeatletter
\providecommand \@ifxundefined [1]{%
 \@ifx{#1\undefined}
}%
\providecommand \@ifnum [1]{%
 \ifnum #1\expandafter \@firstoftwo
 \else \expandafter \@secondoftwo
 \fi
}%
\providecommand \@ifx [1]{%
 \ifx #1\expandafter \@firstoftwo
 \else \expandafter \@secondoftwo
 \fi
}%
\providecommand \natexlab [1]{#1}%
\providecommand \enquote  [1]{``#1''}%
\providecommand \bibnamefont  [1]{#1}%
\providecommand \bibfnamefont [1]{#1}%
\providecommand \citenamefont [1]{#1}%
\providecommand \href@noop [0]{\@secondoftwo}%
\providecommand \href [0]{\begingroup \@sanitize@url \@href}%
\providecommand \@href[1]{\@@startlink{#1}\@@href}%
\providecommand \@@href[1]{\endgroup#1\@@endlink}%
\providecommand \@sanitize@url [0]{\catcode `\\12\catcode `\$12\catcode `\&12\catcode `\#12\catcode `\^12\catcode `\_12\catcode `\%12\relax}%
\providecommand \@@startlink[1]{}%
\providecommand \@@endlink[0]{}%
\providecommand \url  [0]{\begingroup\@sanitize@url \@url }%
\providecommand \@url [1]{\endgroup\@href {#1}{\urlprefix }}%
\providecommand \urlprefix  [0]{URL }%
\providecommand \Eprint [0]{\href }%
\providecommand \doibase [0]{https://doi.org/}%
\providecommand \selectlanguage [0]{\@gobble}%
\providecommand \bibinfo  [0]{\@secondoftwo}%
\providecommand \bibfield  [0]{\@secondoftwo}%
\providecommand \translation [1]{[#1]}%
\providecommand \BibitemOpen [0]{}%
\providecommand \bibitemStop [0]{}%
\providecommand \bibitemNoStop [0]{.\EOS\space}%
\providecommand \EOS [0]{\spacefactor3000\relax}%
\providecommand \BibitemShut  [1]{\csname bibitem#1\endcsname}%
\let\auto@bib@innerbib\@empty
\bibitem [{\citenamefont {Ignatenko}\ \emph {et~al.}(2019)\citenamefont {Ignatenko}, \citenamefont {Buyadzhi}, \citenamefont {Buyadzhi}, \citenamefont {Kuznetsova}, \citenamefont {Mashkantsev},\ and\ \citenamefont {Ternovsky}}]{ignatenko2019nonlinear}%
  \BibitemOpen
  \bibfield  {author} {\bibinfo {author} {\bibfnamefont {A.}~\bibnamefont {Ignatenko}}, \bibinfo {author} {\bibfnamefont {A.}~\bibnamefont {Buyadzhi}}, \bibinfo {author} {\bibfnamefont {V.}~\bibnamefont {Buyadzhi}}, \bibinfo {author} {\bibfnamefont {A.}~\bibnamefont {Kuznetsova}}, \bibinfo {author} {\bibfnamefont {A.}~\bibnamefont {Mashkantsev}},\ and\ \bibinfo {author} {\bibfnamefont {E.}~\bibnamefont {Ternovsky}},\ }\bibfield  {title} {\bibinfo {title} {Nonlinear chaotic dynamics of quantum systems: molecules in an electromagnetic field},\ }in\ \href {https://doi.org/10.1016/bs.aiq.2018.06.006} {\emph {\bibinfo {booktitle} {Adv. Quantum Chem.}}},\ Vol.~\bibinfo {volume} {78}\ (\bibinfo {year} {2019})\ pp.\ \bibinfo {pages} {149--170}\BibitemShut {NoStop}%
\bibitem [{\citenamefont {Melkikh}(2015)}]{melkikh2015nonlinearity}%
  \BibitemOpen
  \bibfield  {author} {\bibinfo {author} {\bibfnamefont {A.~V.}\ \bibnamefont {Melkikh}},\ }\bibfield  {title} {\bibinfo {title} {Nonlinearity of quantum mechanics and solution of the problem of wave function collapse},\ }\href {https://doi.org/10.1088/0253-6102/64/1/47} {\bibfield  {journal} {\bibinfo  {journal} {Commun. Theor. Phys.}\ }\textbf {\bibinfo {volume} {64}},\ \bibinfo {pages} {47} (\bibinfo {year} {2015})}\BibitemShut {NoStop}%
\bibitem [{\citenamefont {Huang}\ \emph {et~al.}(2020)\citenamefont {Huang}, \citenamefont {Zhao}, \citenamefont {Zeng}, \citenamefont {Crunteanu}, \citenamefont {Shum},\ and\ \citenamefont {Yu}}]{huang2020planar}%
  \BibitemOpen
  \bibfield  {author} {\bibinfo {author} {\bibfnamefont {T.}~\bibnamefont {Huang}}, \bibinfo {author} {\bibfnamefont {X.}~\bibnamefont {Zhao}}, \bibinfo {author} {\bibfnamefont {S.}~\bibnamefont {Zeng}}, \bibinfo {author} {\bibfnamefont {A.}~\bibnamefont {Crunteanu}}, \bibinfo {author} {\bibfnamefont {P.~P.}\ \bibnamefont {Shum}},\ and\ \bibinfo {author} {\bibfnamefont {N.}~\bibnamefont {Yu}},\ }\bibfield  {title} {\bibinfo {title} {Planar nonlinear metasurface optics and their applications},\ }\href {https://doi.org/10.1088/1361-6633/abb56e} {\bibfield  {journal} {\bibinfo  {journal} {Rep. Prog. Phys.}\ }\textbf {\bibinfo {volume} {83}},\ \bibinfo {pages} {126101} (\bibinfo {year} {2020})}\BibitemShut {NoStop}%
\bibitem [{\citenamefont {De~Alba}\ \emph {et~al.}(2016)\citenamefont {De~Alba}, \citenamefont {Massel}, \citenamefont {Storch}, \citenamefont {Abhilash}, \citenamefont {Hui}, \citenamefont {McEuen}, \citenamefont {Craighead},\ and\ \citenamefont {Parpia}}]{de2016tunable}%
  \BibitemOpen
  \bibfield  {author} {\bibinfo {author} {\bibfnamefont {R.}~\bibnamefont {De~Alba}}, \bibinfo {author} {\bibfnamefont {F.}~\bibnamefont {Massel}}, \bibinfo {author} {\bibfnamefont {I.~R.}\ \bibnamefont {Storch}}, \bibinfo {author} {\bibfnamefont {T.~S.}\ \bibnamefont {Abhilash}}, \bibinfo {author} {\bibfnamefont {A.}~\bibnamefont {Hui}}, \bibinfo {author} {\bibfnamefont {P.~L.}\ \bibnamefont {McEuen}}, \bibinfo {author} {\bibfnamefont {H.~G.}\ \bibnamefont {Craighead}},\ and\ \bibinfo {author} {\bibfnamefont {J.~M.}\ \bibnamefont {Parpia}},\ }\bibfield  {title} {\bibinfo {title} {Tunable phonon-cavity coupling in graphene membranes},\ }\href {https://doi.org/10.1038/nnano.2016.86} {\bibfield  {journal} {\bibinfo  {journal} {Nat. Nanotechnol.}\ }\textbf {\bibinfo {volume} {11}},\ \bibinfo {pages} {741} (\bibinfo {year} {2016})}\BibitemShut {NoStop}%
\bibitem [{\citenamefont {Rega}(2022)}]{rega2022nonlinear}%
  \BibitemOpen
  \bibfield  {author} {\bibinfo {author} {\bibfnamefont {G.}~\bibnamefont {Rega}},\ }\bibfield  {title} {\bibinfo {title} {Nonlinear dynamics in mechanics: state of the art and expected future developments},\ }\href {https://doi.org/10.1115/1.4054112} {\bibfield  {journal} {\bibinfo  {journal} {J. Comput. Nonlinear Dyn.}\ }\textbf {\bibinfo {volume} {17}},\ \bibinfo {pages} {080802} (\bibinfo {year} {2022})}\BibitemShut {NoStop}%
\bibitem [{\citenamefont {Craighead}(2000)}]{craighead2000nanoelectromechanical}%
  \BibitemOpen
  \bibfield  {author} {\bibinfo {author} {\bibfnamefont {H.~G.}\ \bibnamefont {Craighead}},\ }\bibfield  {title} {\bibinfo {title} {Nanoelectromechanical systems},\ }\href {https://doi.org/10.1126/science.290.5496.1532} {\bibfield  {journal} {\bibinfo  {journal} {Science}\ }\textbf {\bibinfo {volume} {290}},\ \bibinfo {pages} {1532} (\bibinfo {year} {2000})}\BibitemShut {NoStop}%
\bibitem [{\citenamefont {Ben}\ \emph {et~al.}(2023)\citenamefont {Ben}, \citenamefont {Fernando}, \citenamefont {Ou}, \citenamefont {Dupré}, \citenamefont {Ollier}, \citenamefont {Hassani}, \citenamefont {Mizuta},\ and\ \citenamefont {Tsuchiya}}]{ben2023characterisation}%
  \BibitemOpen
  \bibfield  {author} {\bibinfo {author} {\bibfnamefont {F.}~\bibnamefont {Ben}}, \bibinfo {author} {\bibfnamefont {J.}~\bibnamefont {Fernando}}, \bibinfo {author} {\bibfnamefont {J.-Y.}\ \bibnamefont {Ou}}, \bibinfo {author} {\bibfnamefont {C.}~\bibnamefont {Dupré}}, \bibinfo {author} {\bibfnamefont {E.}~\bibnamefont {Ollier}}, \bibinfo {author} {\bibfnamefont {F.~A.}\ \bibnamefont {Hassani}}, \bibinfo {author} {\bibfnamefont {H.}~\bibnamefont {Mizuta}},\ and\ \bibinfo {author} {\bibfnamefont {Y.}~\bibnamefont {Tsuchiya}},\ }\bibfield  {title} {\bibinfo {title} {Characterisation and modelling of nonlinear resonance behaviour on very-high-frequency silicon nanoelectromechanical resonators},\ }\href {https://doi.org/10.1016/j.mne.2023.100212} {\bibfield  {journal} {\bibinfo  {journal} {Micro Nano Eng.}\ }\textbf {\bibinfo {volume} {19}},\ \bibinfo {pages} {100212} (\bibinfo {year} {2023})}\BibitemShut {NoStop}%
\bibitem [{\citenamefont {Asadi}\ \emph {et~al.}(2018)\citenamefont {Asadi}, \citenamefont {Yu},\ and\ \citenamefont {Cho}}]{asadi2018nonlinear}%
  \BibitemOpen
  \bibfield  {author} {\bibinfo {author} {\bibfnamefont {K.}~\bibnamefont {Asadi}}, \bibinfo {author} {\bibfnamefont {J.}~\bibnamefont {Yu}},\ and\ \bibinfo {author} {\bibfnamefont {H.}~\bibnamefont {Cho}},\ }\bibfield  {title} {\bibinfo {title} {Nonlinear couplings and energy transfers in micro-and nano-mechanical resonators: intermodal coupling, internal resonance and synchronization},\ }\href {https://doi.org/10.1098/rsta.2017.0141} {\bibfield  {journal} {\bibinfo  {journal} {Philos. Trans. R. Soc. A}\ }\textbf {\bibinfo {volume} {376}},\ \bibinfo {pages} {20170141} (\bibinfo {year} {2018})}\BibitemShut {NoStop}%
\bibitem [{\citenamefont {Dinh}\ \emph {et~al.}(2023)\citenamefont {Dinh}, \citenamefont {Phan}, \citenamefont {Nguyen}, \citenamefont {Rais-Zadeh}, \citenamefont {Nguyen}, \citenamefont {Song}, \citenamefont {Bell}, \citenamefont {Dao},\ and\ \citenamefont {Deo}}]{dinh2023}%
  \BibitemOpen
  \bibfield  {author} {\bibinfo {author} {\bibfnamefont {T.}~\bibnamefont {Dinh}}, \bibinfo {author} {\bibfnamefont {H.}~\bibnamefont {Phan}}, \bibinfo {author} {\bibfnamefont {N.}~\bibnamefont {Nguyen}}, \bibinfo {author} {\bibfnamefont {M.}~\bibnamefont {Rais-Zadeh}}, \bibinfo {author} {\bibfnamefont {T.}~\bibnamefont {Nguyen}}, \bibinfo {author} {\bibfnamefont {P.}~\bibnamefont {Song}}, \bibinfo {author} {\bibfnamefont {J.}~\bibnamefont {Bell}}, \bibinfo {author} {\bibfnamefont {D.}~\bibnamefont {Dao}},\ and\ \bibinfo {author} {\bibfnamefont {R.}~\bibnamefont {Deo}},\ }\bibfield  {title} {\bibinfo {title} {Micromachined mechanical resonant sensors: From materials, structural designs to applications},\ }\bibfield  {journal} {\bibinfo  {journal} {Adv. Mater. Technol.}\ }\href {https://doi.org/10.1002/admt.202300913} {10.1002/admt.202300913} (\bibinfo {year} {2023})\BibitemShut {NoStop}%
\bibitem [{\citenamefont {Rhoads}\ \emph {et~al.}(2010)\citenamefont {Rhoads}, \citenamefont {Shaw},\ and\ \citenamefont {Turner}}]{rhoads2010nonlinear}%
  \BibitemOpen
  \bibfield  {author} {\bibinfo {author} {\bibfnamefont {J.~F.}\ \bibnamefont {Rhoads}}, \bibinfo {author} {\bibfnamefont {S.~W.}\ \bibnamefont {Shaw}},\ and\ \bibinfo {author} {\bibfnamefont {K.~L.}\ \bibnamefont {Turner}},\ }\bibfield  {title} {\bibinfo {title} {Nonlinear dynamics and its applications in micro-and nanoresonators}\ }\href {https://doi.org/10.1115/1.4001333} {10.1115/1.4001333} (\bibinfo {year} {2010})\BibitemShut {NoStop}%
\bibitem [{\citenamefont {Ekinci}\ and\ \citenamefont {Roukes}(2005)}]{ekinci2005nanoelectromechanical}%
  \BibitemOpen
  \bibfield  {author} {\bibinfo {author} {\bibfnamefont {K.~L.}\ \bibnamefont {Ekinci}}\ and\ \bibinfo {author} {\bibfnamefont {M.~L.}\ \bibnamefont {Roukes}},\ }\bibfield  {title} {\bibinfo {title} {Nanoelectromechanical systems},\ }\href {https://doi.org/10.1063/1.1927327} {\bibfield  {journal} {\bibinfo  {journal} {Rev. Sci. Instrum.}\ }\textbf {\bibinfo {volume} {76}},\ \bibinfo {pages} {061101} (\bibinfo {year} {2005})}\BibitemShut {NoStop}%
\bibitem [{\citenamefont {Wei}\ \emph {et~al.}(2021)\citenamefont {Wei}, \citenamefont {Kuai}, \citenamefont {Bao}, \citenamefont {Wei}, \citenamefont {Yang}, \citenamefont {Song}, \citenamefont {Zhang}, \citenamefont {Yang},\ and\ \citenamefont {Wang}}]{wei2021recent}%
  \BibitemOpen
  \bibfield  {author} {\bibinfo {author} {\bibfnamefont {L.}~\bibnamefont {Wei}}, \bibinfo {author} {\bibfnamefont {X.}~\bibnamefont {Kuai}}, \bibinfo {author} {\bibfnamefont {Y.}~\bibnamefont {Bao}}, \bibinfo {author} {\bibfnamefont {J.}~\bibnamefont {Wei}}, \bibinfo {author} {\bibfnamefont {L.}~\bibnamefont {Yang}}, \bibinfo {author} {\bibfnamefont {P.}~\bibnamefont {Song}}, \bibinfo {author} {\bibfnamefont {M.}~\bibnamefont {Zhang}}, \bibinfo {author} {\bibfnamefont {F.}~\bibnamefont {Yang}},\ and\ \bibinfo {author} {\bibfnamefont {X.}~\bibnamefont {Wang}},\ }\bibfield  {title} {\bibinfo {title} {The recent progress of mems/nems resonators},\ }\href {https://doi.org/10.3390/mi12060724} {\bibfield  {journal} {\bibinfo  {journal} {Micromachines}\ }\textbf {\bibinfo {volume} {12}},\ \bibinfo {pages} {724} (\bibinfo {year} {2021})}\BibitemShut {NoStop}%
\bibitem [{\citenamefont {Samanta}\ \emph {et~al.}(2015)\citenamefont {Samanta}, \citenamefont {Gangavarapu},\ and\ \citenamefont {Naik}}]{samanta2015nonlinear}%
  \BibitemOpen
  \bibfield  {author} {\bibinfo {author} {\bibfnamefont {C.}~\bibnamefont {Samanta}}, \bibinfo {author} {\bibfnamefont {P.~R.~Y.}\ \bibnamefont {Gangavarapu}},\ and\ \bibinfo {author} {\bibfnamefont {A.~K.}\ \bibnamefont {Naik}},\ }\bibfield  {title} {\bibinfo {title} {Nonlinear mode coupling and internal resonances in mos$_2$ nanoelectromechanical system},\ }\href {https://doi.org/10.1063/1.4934708} {\bibfield  {journal} {\bibinfo  {journal} {Appl. Phys. Lett.}\ }\textbf {\bibinfo {volume} {107}},\ \bibinfo {pages} {173101} (\bibinfo {year} {2015})}\BibitemShut {NoStop}%
\bibitem [{\citenamefont {Rieger}\ \emph {et~al.}(2012)\citenamefont {Rieger}, \citenamefont {Faust}, \citenamefont {Seitner}, \citenamefont {Kotthaus},\ and\ \citenamefont {Weig}}]{Rieger2012}%
  \BibitemOpen
  \bibfield  {author} {\bibinfo {author} {\bibfnamefont {J.}~\bibnamefont {Rieger}}, \bibinfo {author} {\bibfnamefont {T.}~\bibnamefont {Faust}}, \bibinfo {author} {\bibfnamefont {M.~J.}\ \bibnamefont {Seitner}}, \bibinfo {author} {\bibfnamefont {J.~P.}\ \bibnamefont {Kotthaus}},\ and\ \bibinfo {author} {\bibfnamefont {E.~M.}\ \bibnamefont {Weig}},\ }\bibfield  {title} {\bibinfo {title} {Frequency and q factor control of nanomechanical resonators},\ }\href {https://doi.org/10.1063/1.4751351} {\bibfield  {journal} {\bibinfo  {journal} {App. Phys. Lett.}\ }\textbf {\bibinfo {volume} {101}},\ \bibinfo {pages} {103110} (\bibinfo {year} {2012})}\BibitemShut {NoStop}%
\bibitem [{\citenamefont {Tu}\ \emph {et~al.}(2020)\citenamefont {Tu}, \citenamefont {Lee},\ and\ \citenamefont {Zhang}}]{Tu2020}%
  \BibitemOpen
  \bibfield  {author} {\bibinfo {author} {\bibfnamefont {C.}~\bibnamefont {Tu}}, \bibinfo {author} {\bibfnamefont {J.~E.-Y.}\ \bibnamefont {Lee}},\ and\ \bibinfo {author} {\bibfnamefont {X.-S.}\ \bibnamefont {Zhang}},\ }\bibfield  {title} {\bibinfo {title} {Dissipation analysis methods and q-enhancement strategies in piezoelectric mems laterally vibrating resonators: A review},\ }\href {https://doi.org/10.3390/s20174978} {\bibfield  {journal} {\bibinfo  {journal} {Sensors}\ }\textbf {\bibinfo {volume} {20}},\ \bibinfo {pages} {4978} (\bibinfo {year} {2020})}\BibitemShut {NoStop}%
\bibitem [{\citenamefont {Jia}\ \emph {et~al.}(2021)\citenamefont {Jia}, \citenamefont {Xu},\ and\ \citenamefont {Li}}]{Jia2021}%
  \BibitemOpen
  \bibfield  {author} {\bibinfo {author} {\bibfnamefont {H.}~\bibnamefont {Jia}}, \bibinfo {author} {\bibfnamefont {P.}~\bibnamefont {Xu}},\ and\ \bibinfo {author} {\bibfnamefont {X.}~\bibnamefont {Li}},\ }\bibfield  {title} {\bibinfo {title} {Integrated resonant micro/nano gravimetric sensors for bio/chemical detection in air and liquid},\ }\href {https://doi.org/10.3390/mi12060645} {\bibfield  {journal} {\bibinfo  {journal} {Micromachines}\ }\textbf {\bibinfo {volume} {12}},\ \bibinfo {pages} {645} (\bibinfo {year} {2021})}\BibitemShut {NoStop}%
\bibitem [{\citenamefont {Braakman}\ and\ \citenamefont {Poggio}(2019)}]{braakman2019force}%
  \BibitemOpen
  \bibfield  {author} {\bibinfo {author} {\bibfnamefont {F.~R.}\ \bibnamefont {Braakman}}\ and\ \bibinfo {author} {\bibfnamefont {M.}~\bibnamefont {Poggio}},\ }\bibfield  {title} {\bibinfo {title} {Force sensing with nanowire cantilevers},\ }\href {https://doi.org/10.1088/1361-6528/ab19cf} {\bibfield  {journal} {\bibinfo  {journal} {Nanotechnology}\ }\textbf {\bibinfo {volume} {30}},\ \bibinfo {pages} {332001} (\bibinfo {year} {2019})}\BibitemShut {NoStop}%
\bibitem [{\citenamefont {Samanta}\ \emph {et~al.}(2018)\citenamefont {Samanta}, \citenamefont {Arora},\ and\ \citenamefont {Naik}}]{samanta2018tuning}%
  \BibitemOpen
  \bibfield  {author} {\bibinfo {author} {\bibfnamefont {C.}~\bibnamefont {Samanta}}, \bibinfo {author} {\bibfnamefont {N.}~\bibnamefont {Arora}},\ and\ \bibinfo {author} {\bibfnamefont {A.~K.}\ \bibnamefont {Naik}},\ }\bibfield  {title} {\bibinfo {title} {Tuning of geometric nonlinearity in ultrathin nanoelectromechanical systems},\ }\href {https://doi.org/10.1063/1.5026775} {\bibfield  {journal} {\bibinfo  {journal} {Appl. Phys. Lett.}\ }\textbf {\bibinfo {volume} {113}},\ \bibinfo {pages} {113101} (\bibinfo {year} {2018})}\BibitemShut {NoStop}%
\bibitem [{\citenamefont {Tang}\ \emph {et~al.}(2005)\citenamefont {Tang}, \citenamefont {Xu}, \citenamefont {Li},\ and\ \citenamefont {Aluru}}]{tang2005physical}%
  \BibitemOpen
  \bibfield  {author} {\bibinfo {author} {\bibfnamefont {Z.}~\bibnamefont {Tang}}, \bibinfo {author} {\bibfnamefont {Y.}~\bibnamefont {Xu}}, \bibinfo {author} {\bibfnamefont {G.}~\bibnamefont {Li}},\ and\ \bibinfo {author} {\bibfnamefont {N.~R.}\ \bibnamefont {Aluru}},\ }\bibfield  {title} {\bibinfo {title} {Physical models for coupled electromechanical analysis of silicon nanoelectromechanical systems},\ }\href {https://doi.org/10.1063/1.1897483} {\bibfield  {journal} {\bibinfo  {journal} {J. Appl. Phys.}\ }\textbf {\bibinfo {volume} {97}},\ \bibinfo {pages} {114304} (\bibinfo {year} {2005})}\BibitemShut {NoStop}%
\bibitem [{\citenamefont {Pavithran}\ \emph {et~al.}(2023)\citenamefont {Pavithran}, \citenamefont {Midhun},\ and\ \citenamefont {Sujith}}]{pavithran2023tipping}%
  \BibitemOpen
  \bibfield  {author} {\bibinfo {author} {\bibfnamefont {I.}~\bibnamefont {Pavithran}}, \bibinfo {author} {\bibfnamefont {P.~R.}\ \bibnamefont {Midhun}},\ and\ \bibinfo {author} {\bibfnamefont {R.~I.}\ \bibnamefont {Sujith}},\ }\bibfield  {title} {\bibinfo {title} {Tipping in complex systems under fast variations of parameters},\ }\href {https://doi.org/10.1063/5.0162503} {\bibfield  {journal} {\bibinfo  {journal} {{Chaos}}\ }\textbf {\bibinfo {volume} {33}},\ \bibinfo {pages} {083123} (\bibinfo {year} {2023})}\BibitemShut {NoStop}%
\bibitem [{\citenamefont {Bury}\ \emph {et~al.}(2020)\citenamefont {Bury}, \citenamefont {Bauch},\ and\ \citenamefont {Anand}}]{bury2020detecting}%
  \BibitemOpen
  \bibfield  {author} {\bibinfo {author} {\bibfnamefont {T.~M.}\ \bibnamefont {Bury}}, \bibinfo {author} {\bibfnamefont {C.~T.}\ \bibnamefont {Bauch}},\ and\ \bibinfo {author} {\bibfnamefont {M.}~\bibnamefont {Anand}},\ }\bibfield  {title} {\bibinfo {title} {Detecting and distinguishing tipping points using spectral early warning signals},\ }\href {https://doi.org/10.1098/rsif.2020.0482} {\bibfield  {journal} {\bibinfo  {journal} {{J. R. Soc. Interface}}\ }\textbf {\bibinfo {volume} {17}},\ \bibinfo {pages} {20200482} (\bibinfo {year} {2020})}\BibitemShut {NoStop}%
\bibitem [{\citenamefont {Scheffer}\ \emph {et~al.}(2009)\citenamefont {Scheffer}, \citenamefont {Bascompte}, \citenamefont {Brock}, \citenamefont {Brovkin}, \citenamefont {Carpenter}, \citenamefont {Dakos}, \citenamefont {Held}, \citenamefont {Van~Nes}, \citenamefont {Rietkerk},\ and\ \citenamefont {Sugihara}}]{Early}%
  \BibitemOpen
  \bibfield  {author} {\bibinfo {author} {\bibfnamefont {M.}~\bibnamefont {Scheffer}}, \bibinfo {author} {\bibfnamefont {J.}~\bibnamefont {Bascompte}}, \bibinfo {author} {\bibfnamefont {W.~A.}\ \bibnamefont {Brock}}, \bibinfo {author} {\bibfnamefont {V.}~\bibnamefont {Brovkin}}, \bibinfo {author} {\bibfnamefont {S.~R.}\ \bibnamefont {Carpenter}}, \bibinfo {author} {\bibfnamefont {V.}~\bibnamefont {Dakos}}, \bibinfo {author} {\bibfnamefont {H.}~\bibnamefont {Held}}, \bibinfo {author} {\bibfnamefont {E.~H.}\ \bibnamefont {Van~Nes}}, \bibinfo {author} {\bibfnamefont {M.}~\bibnamefont {Rietkerk}},\ and\ \bibinfo {author} {\bibfnamefont {G.}~\bibnamefont {Sugihara}},\ }\bibfield  {title} {\bibinfo {title} {Early-warning signals for critical transitions},\ }\href {https://doi.org/10.1038/nature08227} {\bibfield  {journal} {\bibinfo  {journal} {Nature}\ }\textbf {\bibinfo {volume} {461}},\ \bibinfo {pages} {53} (\bibinfo {year} {2009})}\BibitemShut {NoStop}%
\bibitem [{\citenamefont {Mosca}\ \emph {et~al.}(2018)\citenamefont {Mosca}, \citenamefont {Ricciardi}, \citenamefont {Hansson}, \citenamefont {Parisi}, \citenamefont {Maddaloni}, \citenamefont {Erkintalo}, \citenamefont {Wabnitz}, \citenamefont {Leo}, \citenamefont {Natale},\ and\ \citenamefont {Rosa}}]{mosca}%
  \BibitemOpen
  \bibfield  {author} {\bibinfo {author} {\bibfnamefont {S.}~\bibnamefont {Mosca}}, \bibinfo {author} {\bibfnamefont {I.}~\bibnamefont {Ricciardi}}, \bibinfo {author} {\bibfnamefont {T.}~\bibnamefont {Hansson}}, \bibinfo {author} {\bibfnamefont {M.}~\bibnamefont {Parisi}}, \bibinfo {author} {\bibfnamefont {P.}~\bibnamefont {Maddaloni}}, \bibinfo {author} {\bibfnamefont {M.}~\bibnamefont {Erkintalo}}, \bibinfo {author} {\bibfnamefont {S.}~\bibnamefont {Wabnitz}}, \bibinfo {author} {\bibfnamefont {F.}~\bibnamefont {Leo}}, \bibinfo {author} {\bibfnamefont {P.~D.}\ \bibnamefont {Natale}},\ and\ \bibinfo {author} {\bibfnamefont {M.~D.}\ \bibnamefont {Rosa}},\ }\bibfield  {title} {\bibinfo {title} {Modulation instability induced frequency comb generation in a continuously pumped optical parametric oscillator},\ }\bibfield  {journal} {\bibinfo  {journal} {Physical Review Letters}\ }\textbf {\bibinfo {volume} {121}},\ \href {https://doi.org/10.1103/PhysRevLett.121.093903} {10.1103/PhysRevLett.121.093903} (\bibinfo
  {year} {2018})\BibitemShut {NoStop}%
\bibitem [{\citenamefont {Czaplewski}\ \emph {et~al.}(2018)\citenamefont {Czaplewski}, \citenamefont {Chen}, \citenamefont {Lopez}, \citenamefont {Shoshani}, \citenamefont {Eriksson}, \citenamefont {Strachan},\ and\ \citenamefont {Shaw}}]{bifurcation}%
  \BibitemOpen
  \bibfield  {author} {\bibinfo {author} {\bibfnamefont {D.~A.}\ \bibnamefont {Czaplewski}}, \bibinfo {author} {\bibfnamefont {C.}~\bibnamefont {Chen}}, \bibinfo {author} {\bibfnamefont {D.}~\bibnamefont {Lopez}}, \bibinfo {author} {\bibfnamefont {O.}~\bibnamefont {Shoshani}}, \bibinfo {author} {\bibfnamefont {A.~M.}\ \bibnamefont {Eriksson}}, \bibinfo {author} {\bibfnamefont {S.}~\bibnamefont {Strachan}},\ and\ \bibinfo {author} {\bibfnamefont {S.~W.}\ \bibnamefont {Shaw}},\ }\bibfield  {title} {\bibinfo {title} {Bifurcation generated mechanical frequency comb},\ }\href {https://doi.org/10.1103/PhysRevLett.121.244302} {\bibfield  {journal} {\bibinfo  {journal} {Physical Review Letters}\ }\textbf {\bibinfo {volume} {121}},\ \bibinfo {pages} {244302} (\bibinfo {year} {2018})}\BibitemShut {NoStop}%
\bibitem [{\citenamefont {Ochs}\ \emph {et~al.}(2022)\citenamefont {Ochs}, \citenamefont {Bone{\ss}}, \citenamefont {Rastelli}, \citenamefont {Seitner}, \citenamefont {Belzig}, \citenamefont {Dykman},\ and\ \citenamefont {Weig}}]{ochs2022frequency}%
  \BibitemOpen
  \bibfield  {author} {\bibinfo {author} {\bibfnamefont {J.~S.}\ \bibnamefont {Ochs}}, \bibinfo {author} {\bibfnamefont {D.~K.~J.}\ \bibnamefont {Bone{\ss}}}, \bibinfo {author} {\bibfnamefont {G.}~\bibnamefont {Rastelli}}, \bibinfo {author} {\bibfnamefont {M.}~\bibnamefont {Seitner}}, \bibinfo {author} {\bibfnamefont {W.}~\bibnamefont {Belzig}}, \bibinfo {author} {\bibfnamefont {M.~I.}\ \bibnamefont {Dykman}},\ and\ \bibinfo {author} {\bibfnamefont {E.~M.}\ \bibnamefont {Weig}},\ }\bibfield  {title} {\bibinfo {title} {Frequency comb from a single driven nonlinear nanomechanical mode},\ }\href {https://doi.org/10.1103/PhysRevX.12.041019} {\bibfield  {journal} {\bibinfo  {journal} {{Phys. Rev. X}}\ }\textbf {\bibinfo {volume} {12}},\ \bibinfo {pages} {041019} (\bibinfo {year} {2022})}\BibitemShut {NoStop}%
\bibitem [{\citenamefont {Rahmanian}\ \emph {et~al.}(2025)\citenamefont {Rahmanian}, \citenamefont {Mouharrar}, \citenamefont {Abdelrahman}, \citenamefont {Akbari}, \citenamefont {Shama}, \citenamefont {Musselman}, \citenamefont {Mu{\~n}oz-Rojas}, \citenamefont {Basrour},\ and\ \citenamefont {Abdel~Rahman}}]{rahmanian2025nems}%
  \BibitemOpen
  \bibfield  {author} {\bibinfo {author} {\bibfnamefont {S.}~\bibnamefont {Rahmanian}}, \bibinfo {author} {\bibfnamefont {H.}~\bibnamefont {Mouharrar}}, \bibinfo {author} {\bibfnamefont {R.}~\bibnamefont {Abdelrahman}}, \bibinfo {author} {\bibfnamefont {M.}~\bibnamefont {Akbari}}, \bibinfo {author} {\bibfnamefont {Y.~S.}\ \bibnamefont {Shama}}, \bibinfo {author} {\bibfnamefont {K.}~\bibnamefont {Musselman}}, \bibinfo {author} {\bibfnamefont {D.}~\bibnamefont {Mu{\~n}oz-Rojas}}, \bibinfo {author} {\bibfnamefont {S.}~\bibnamefont {Basrour}},\ and\ \bibinfo {author} {\bibfnamefont {E.}~\bibnamefont {Abdel~Rahman}},\ }\bibfield  {title} {\bibinfo {title} {Nems generated electromechanical frequency combs},\ }\href {https://doi.org/10.1038/s41378-024-00860-9} {\bibfield  {journal} {\bibinfo  {journal} {{Microsyst. Nanoeng.}}\ }\textbf {\bibinfo {volume} {11}},\ \bibinfo {pages} {8} (\bibinfo {year} {2025})}\BibitemShut {NoStop}%
\bibitem [{\citenamefont {Park}\ and\ \citenamefont {Ansari}(2019)}]{park2019formation}%
  \BibitemOpen
  \bibfield  {author} {\bibinfo {author} {\bibfnamefont {M.}~\bibnamefont {Park}}\ and\ \bibinfo {author} {\bibfnamefont {A.}~\bibnamefont {Ansari}},\ }\bibfield  {title} {\bibinfo {title} {Formation, evolution, and tuning of frequency combs in microelectromechanical resonators},\ }\href {https://doi.org/10.1109/JMEMS.2019.2898003} {\bibfield  {journal} {\bibinfo  {journal} {{J. Microelectromech. Syst.}}\ }\textbf {\bibinfo {volume} {28}},\ \bibinfo {pages} {429} (\bibinfo {year} {2019})}\BibitemShut {NoStop}%
\bibitem [{\citenamefont {Chiout}\ \emph {et~al.}(2021)\citenamefont {Chiout}, \citenamefont {Correia}, \citenamefont {Zhao}, \citenamefont {Johnson}, \citenamefont {Pierucci}, \citenamefont {Oehler}, \citenamefont {Ouerghi},\ and\ \citenamefont {Chaste}}]{chiout2021multi}%
  \BibitemOpen
  \bibfield  {author} {\bibinfo {author} {\bibfnamefont {A.}~\bibnamefont {Chiout}}, \bibinfo {author} {\bibfnamefont {F.}~\bibnamefont {Correia}}, \bibinfo {author} {\bibfnamefont {M.-Q.}\ \bibnamefont {Zhao}}, \bibinfo {author} {\bibfnamefont {A.~T.}\ \bibnamefont {Johnson}}, \bibinfo {author} {\bibfnamefont {D.}~\bibnamefont {Pierucci}}, \bibinfo {author} {\bibfnamefont {F.}~\bibnamefont {Oehler}}, \bibinfo {author} {\bibfnamefont {A.}~\bibnamefont {Ouerghi}},\ and\ \bibinfo {author} {\bibfnamefont {J.}~\bibnamefont {Chaste}},\ }\bibfield  {title} {\bibinfo {title} {Multi-order phononic frequency comb generation within a mos2 electromechanical resonator},\ }\bibfield  {journal} {\bibinfo  {journal} {Appl. Phys. Lett.}\ }\href {https://doi.org/10.1063/5.0059015} {10.1063/5.0059015} (\bibinfo {year} {2021})\BibitemShut {NoStop}%
\bibitem [{\citenamefont {Wu}\ \emph {et~al.}(2025)\citenamefont {Wu}, \citenamefont {Song}, \citenamefont {Zang}, \citenamefont {Li}, \citenamefont {Zhang},\ and\ \citenamefont {Shao}}]{wu2025limit}%
  \BibitemOpen
  \bibfield  {author} {\bibinfo {author} {\bibfnamefont {J.}~\bibnamefont {Wu}}, \bibinfo {author} {\bibfnamefont {P.}~\bibnamefont {Song}}, \bibinfo {author} {\bibfnamefont {S.}~\bibnamefont {Zang}}, \bibinfo {author} {\bibfnamefont {L.}~\bibnamefont {Li}}, \bibinfo {author} {\bibfnamefont {W.}~\bibnamefont {Zhang}},\ and\ \bibinfo {author} {\bibfnamefont {L.}~\bibnamefont {Shao}},\ }\bibfield  {title} {\bibinfo {title} {Limit cycle convergence leads to period-doubling and cyclic-fold bifurcation in internal resonance-induced mechanical frequency combs},\ }\bibfield  {journal} {\bibinfo  {journal} {Nonlinear Dyn.}\ }\href {https://doi.org/10.1007/s11071-025-11169-1} {10.1007/s11071-025-11169-1} (\bibinfo {year} {2025})\BibitemShut {NoStop}%
\bibitem [{\citenamefont {Seitner}\ \emph {et~al.}(2017)\citenamefont {Seitner}, \citenamefont {Abdi}, \citenamefont {Ridolfo}, \citenamefont {Hartmann},\ and\ \citenamefont {Weig}}]{Seitner2017}%
  \BibitemOpen
  \bibfield  {author} {\bibinfo {author} {\bibfnamefont {M.~J.}\ \bibnamefont {Seitner}}, \bibinfo {author} {\bibfnamefont {M.}~\bibnamefont {Abdi}}, \bibinfo {author} {\bibfnamefont {A.}~\bibnamefont {Ridolfo}}, \bibinfo {author} {\bibfnamefont {M.~J.}\ \bibnamefont {Hartmann}},\ and\ \bibinfo {author} {\bibfnamefont {E.~M.}\ \bibnamefont {Weig}},\ }\bibfield  {title} {\bibinfo {title} {Parametric oscillation, frequency mixing, and injection locking of strongly coupled nanomechanical resonator modes},\ }\href {https://doi.org/10.1103/PhysRevLett.118.254301} {\bibfield  {journal} {\bibinfo  {journal} {Phys. Rev. Lett.}\ }\textbf {\bibinfo {volume} {118}},\ \bibinfo {pages} {254301} (\bibinfo {year} {2017})}\BibitemShut {NoStop}%
\bibitem [{\citenamefont {Faust}\ \emph {et~al.}(2012)\citenamefont {Faust}, \citenamefont {Rieger}, \citenamefont {Seitner}, \citenamefont {Krenn}, \citenamefont {Kotthaus},\ and\ \citenamefont {Weig}}]{Faust2012}%
  \BibitemOpen
  \bibfield  {author} {\bibinfo {author} {\bibfnamefont {T.}~\bibnamefont {Faust}}, \bibinfo {author} {\bibfnamefont {J.}~\bibnamefont {Rieger}}, \bibinfo {author} {\bibfnamefont {M.~J.}\ \bibnamefont {Seitner}}, \bibinfo {author} {\bibfnamefont {P.}~\bibnamefont {Krenn}}, \bibinfo {author} {\bibfnamefont {J.~P.}\ \bibnamefont {Kotthaus}},\ and\ \bibinfo {author} {\bibfnamefont {E.~M.}\ \bibnamefont {Weig}},\ }\bibfield  {title} {\bibinfo {title} {Nonadiabatic dynamics of two strongly coupled nanomechanical resonator modes},\ }\href {https://doi.org/10.1103/PhysRevLett.109.037205} {\bibfield  {journal} {\bibinfo  {journal} {Phys. Rev. Lett.}\ }\textbf {\bibinfo {volume} {109}},\ \bibinfo {pages} {037205} (\bibinfo {year} {2012})}\BibitemShut {NoStop}%
\bibitem [{\citenamefont {Barakat}\ \emph {et~al.}(2024)\citenamefont {Barakat}, \citenamefont {Chowdhury}, \citenamefont {Le},\ and\ \citenamefont {Weig}}]{EvaLast}%
  \BibitemOpen
  \bibfield  {author} {\bibinfo {author} {\bibfnamefont {A.~A.}\ \bibnamefont {Barakat}}, \bibinfo {author} {\bibfnamefont {A.}~\bibnamefont {Chowdhury}}, \bibinfo {author} {\bibfnamefont {A.~T.}\ \bibnamefont {Le}},\ and\ \bibinfo {author} {\bibfnamefont {E.~M.}\ \bibnamefont {Weig}},\ }\bibfield  {title} {\bibinfo {title} {Modal coupling impacts the parametric normal mode splitting: Quantifying the tunable mode coupling of a nanomechanical resonator},\ }\bibfield  {journal} {\bibinfo  {journal} {arXiv}\ }\href {https://doi.org/10.48550/arXiv.2412.16767} {10.48550/arXiv.2412.16767} (\bibinfo {year} {2024})\BibitemShut {NoStop}%
\bibitem [{\citenamefont {Strogatz}(2014)}]{chaosbook}%
  \BibitemOpen
  \bibfield  {author} {\bibinfo {author} {\bibfnamefont {S.~H.}\ \bibnamefont {Strogatz}},\ }\href {https://doi.org/10.1201/9780429492563} {\emph {\bibinfo {title} {Nonlinear Dynamics and Chaos: With Applications to Physics, Biology, Chemistry, and Engineering}}},\ \bibinfo {edition} {2nd}\ ed.,\ Studies in Nonlinearity\ (\bibinfo  {publisher} {Westview Press},\ \bibinfo {year} {2014})\BibitemShut {NoStop}%
\bibitem [{\citenamefont {Shelton}\ \emph {et~al.}(2011)\citenamefont {Shelton}, \citenamefont {Boreman}, \citenamefont {Peters}, \citenamefont {Brener}, \citenamefont {Sinclair}, \citenamefont {Ginn},\ and\ \citenamefont {Coffey}}]{shelton2011strong}%
  \BibitemOpen
  \bibfield  {author} {\bibinfo {author} {\bibfnamefont {D.~J.}\ \bibnamefont {Shelton}}, \bibinfo {author} {\bibfnamefont {G.~D.}\ \bibnamefont {Boreman}}, \bibinfo {author} {\bibfnamefont {D.~W.}\ \bibnamefont {Peters}}, \bibinfo {author} {\bibfnamefont {I.}~\bibnamefont {Brener}}, \bibinfo {author} {\bibfnamefont {M.~B.}\ \bibnamefont {Sinclair}}, \bibinfo {author} {\bibfnamefont {J.~C.}\ \bibnamefont {Ginn}},\ and\ \bibinfo {author} {\bibfnamefont {K.~R.}\ \bibnamefont {Coffey}},\ }\bibfield  {title} {\bibinfo {title} {Strong coupling between nanoscale metamaterials and phonons},\ }\href {https://doi.org/10.1021/nl200689z} {\bibfield  {journal} {\bibinfo  {journal} {Nano Letters}\ }\textbf {\bibinfo {volume} {11}},\ \bibinfo {pages} {2104} (\bibinfo {year} {2011})}\BibitemShut {NoStop}%
\bibitem [{\citenamefont {Kashinath}\ \emph {et~al.}(2014)\citenamefont {Kashinath}, \citenamefont {Juniper},\ and\ \citenamefont {Waugh}}]{kashinath2014nonlinear}%
  \BibitemOpen
  \bibfield  {author} {\bibinfo {author} {\bibfnamefont {K.}~\bibnamefont {Kashinath}}, \bibinfo {author} {\bibfnamefont {M.~P.}\ \bibnamefont {Juniper}},\ and\ \bibinfo {author} {\bibfnamefont {I.~C.}\ \bibnamefont {Waugh}},\ }\bibfield  {title} {\bibinfo {title} {Nonlinear self-excited thermoacoustic oscillations of a ducted premixed flame: Bifurcations and routes to chaos},\ }\href {https://doi.org/10.1017/jfm.2014.601} {\bibfield  {journal} {\bibinfo  {journal} {Journal of Fluid Mechanics}\ }\textbf {\bibinfo {volume} {761}},\ \bibinfo {pages} {399} (\bibinfo {year} {2014})}\BibitemShut {NoStop}%
\bibitem [{\citenamefont {Poularikas}\ and\ \citenamefont {Grigoryan}(2018)}]{hilbert}%
  \BibitemOpen
  \bibfield  {author} {\bibinfo {author} {\bibfnamefont {A.~D.}\ \bibnamefont {Poularikas}}\ and\ \bibinfo {author} {\bibfnamefont {A.~M.}\ \bibnamefont {Grigoryan}},\ }\href {https://doi.org/10.1201/9781315218915} {\emph {\bibinfo {title} {Transforms and Applications Handbook}}}\ (\bibinfo  {publisher} {CRC Press},\ \bibinfo {year} {2018})\BibitemShut {NoStop}%
\bibitem [{\citenamefont {Mahdavi}\ \emph {et~al.}(2025)\citenamefont {Mahdavi}, \citenamefont {Zarei},\ and\ \citenamefont {Shahbazi}}]{mahdavi2025synchronization}%
  \BibitemOpen
  \bibfield  {author} {\bibinfo {author} {\bibfnamefont {E.}~\bibnamefont {Mahdavi}}, \bibinfo {author} {\bibfnamefont {M.}~\bibnamefont {Zarei}},\ and\ \bibinfo {author} {\bibfnamefont {F.}~\bibnamefont {Shahbazi}},\ }\bibfield  {title} {\bibinfo {title} {Synchronization of two coupled massive oscillators in the time-delayed kuramoto model},\ }\href {https://doi.org/10.1063/5.0228203} {\bibfield  {journal} {\bibinfo  {journal} {Chaos}\ }\textbf {\bibinfo {volume} {35}},\ \bibinfo {pages} {013109} (\bibinfo {year} {2025})}\BibitemShut {NoStop}%
\bibitem [{\citenamefont {Koike}\ \emph {et~al.}(2024)\citenamefont {Koike}, \citenamefont {Takayasu},\ and\ \citenamefont {Takayasu}}]{koike2024bifurcation}%
  \BibitemOpen
  \bibfield  {author} {\bibinfo {author} {\bibfnamefont {H.}~\bibnamefont {Koike}}, \bibinfo {author} {\bibfnamefont {H.}~\bibnamefont {Takayasu}},\ and\ \bibinfo {author} {\bibfnamefont {M.}~\bibnamefont {Takayasu}},\ }\bibfield  {title} {\bibinfo {title} {Bifurcation and hysteresis in a nonlinear transport model on network motifs},\ }\href {https://doi.org/10.1103/PhysRevResearch.6.013059} {\bibfield  {journal} {\bibinfo  {journal} {Physical Review Research}\ }\textbf {\bibinfo {volume} {6}},\ \bibinfo {pages} {013059} (\bibinfo {year} {2024})}\BibitemShut {NoStop}%
\bibitem [{\citenamefont {Allemeier}\ \emph {et~al.}(2025)\citenamefont {Allemeier}, \citenamefont {Kaya}, \citenamefont {Hanay},\ and\ \citenamefont {Ekinci}}]{Allemeier25}%
  \BibitemOpen
  \bibfield  {author} {\bibinfo {author} {\bibfnamefont {D.}~\bibnamefont {Allemeier}}, \bibinfo {author} {\bibfnamefont {I.~I.}\ \bibnamefont {Kaya}}, \bibinfo {author} {\bibfnamefont {M.~S.}\ \bibnamefont {Hanay}},\ and\ \bibinfo {author} {\bibfnamefont {K.~L.}\ \bibnamefont {Ekinci}},\ }\bibfield  {title} {\bibinfo {title} {Multistability and noise-induced transitions in dispersively coupled nonlinear nanomechanical modes},\ }\href {https://doi.org/10.1103/dtlz-pmnr} {\bibfield  {journal} {\bibinfo  {journal} {Phys. Rev. A}\ }\textbf {\bibinfo {volume} {112}},\ \bibinfo {pages} {023505} (\bibinfo {year} {2025})}\BibitemShut {NoStop}%
\bibitem [{\citenamefont {Pisarchik}\ and\ \citenamefont {Feudel}(2014)}]{pisarchik2014control}%
  \BibitemOpen
  \bibfield  {author} {\bibinfo {author} {\bibfnamefont {A.~N.}\ \bibnamefont {Pisarchik}}\ and\ \bibinfo {author} {\bibfnamefont {U.}~\bibnamefont {Feudel}},\ }\bibfield  {title} {\bibinfo {title} {Control of multistability},\ }\href {https://doi.org/10.1016/j.physrep.2014.02.007} {\bibfield  {journal} {\bibinfo  {journal} {Physics Reports}\ }\textbf {\bibinfo {volume} {540}},\ \bibinfo {pages} {167} (\bibinfo {year} {2014})}\BibitemShut {NoStop}%
\bibitem [{\citenamefont {Phillips}\ \emph {et~al.}(2025)\citenamefont {Phillips}, \citenamefont {Lindner},\ and\ \citenamefont {Kantz}}]{phillips2025stabilizing}%
  \BibitemOpen
  \bibfield  {author} {\bibinfo {author} {\bibfnamefont {E.~T.}\ \bibnamefont {Phillips}}, \bibinfo {author} {\bibfnamefont {B.}~\bibnamefont {Lindner}},\ and\ \bibinfo {author} {\bibfnamefont {H.}~\bibnamefont {Kantz}},\ }\bibfield  {title} {\bibinfo {title} {Stabilizing role of multiplicative noise in nonconfining potentials},\ }\href {https://doi.org/10.1103/PhysRevResearch.7.023146} {\bibfield  {journal} {\bibinfo  {journal} {Physical Review Research}\ }\textbf {\bibinfo {volume} {7}},\ \bibinfo {pages} {023146} (\bibinfo {year} {2025})}\BibitemShut {NoStop}%
\end{thebibliography}

\end{document}